# Waveform cross correlation applied to earthquakes in the Atlantic Ocean

Dmitry Bobrov, Ivan Kitov, and Mikhail Rozhkov

Comprehensive Nuclear-Test-Ban Treaty Organization

Abstract

We assess the level of cross correlation between P-waves generated by earthquakes in the Atlantic Ocean and measured by 22 array stations of the International Monitoring System (IMS). There are 931 events with 6,411 arrivals in 2011 and 2012. Station TORD was the most sensitive and detected 868 from 931 events. We constructed several 931×931 matrices of cross correlation coefficients (CCs) for individual stations and also for average and cumulative CCs. These matrices characterize the detection performance of the involved stations and the IMS. Sixty earthquakes located in the northern hemisphere were selected as master events for signal detection and building of events populating a cross correlation Standard Event List (XSEL) for the first halves of 2009 and 2012. High-quality signals (SNR>5.0) recorded by 10 most sensitive stations were used as waveform templates. In order to quantitatively estimate the gain in the completeness and resolution of the XSEL we compared it with the Reviewed Event Bulletin (REB) of the International Data Centre (IDC) for the North Atlantic (NA) and with the ISC Bulletin. Machine learning and classification algorithms were successfully applied to automatically reject invalid events in the XSEL for 2009.

# Introduction

The International Monitoring System of the (currently Provisional) Technical Secretariat (TS) of the Comprehensive Nuclear-Test-Ban Treaty Organization (CTBTO) is a global network. By design, the IMS includes 50 primary and 120 auxiliary seismic stations. The International Data Centre (IDC) of the CTBTO receives, collects, processes, analyses, reports on and archives data from the IMS. The IDC is responsible for automatic and interactive processing of the IMS data and for standard IDC products. The Reviewed Event Bulletin (REB) is the principal product of the IDC. Its quality is guaranteed by multistage automatic processing and strict rules of interactive review conducted by experienced analysts. The REB is available to the seismological community as a bulletin distributed by the International Seismological Centre. The IDC is one of major contributors of the ISC. Because the IMS is designed to have globally uniform coverage many events in the ISC bulletins are unique to the IDC. Thus, any improvement in the completeness of the IDC bulletin is immediately reflected in the quality of the ISC bulletin representing one of the main sources of seismological information for numerous scientific studies.

The input of the IDC is especially important in the zones not covered by regional networks. The Atlantic Ocean (seismic region 32 in the Flinn-Engdahl regionalization scheme) is a good example. Woessner and Wiemer (2005) estimated the magnitude of completeness of the ISC catalogue from 1980 to 2001 between 4.3 in the northernmost part of the Atlantic Ocean, 4.7 in the central segment, and 5.0 in the southernmost part. For the IDC, Kitov *et al*. (2011) and Bobrov *et al*. (2011) found practically the same value for the North Atlantic and a slightly lower threshold for the South Atlantic likely associated with the difference in body wave magnitude definitions used by the ISC and IDC as described by Kim *et al*. (2001) and Coyne *et al*. (2012). In this paper, we assess the possibility to reduce the completeness threshold by 0.3 to 0.4 units of magnitude for the whole Atlantic Ocean using waveform cross correlation with master templates carefully selected from the IDC seismic archive.

The improvement in detection and phase association is the principal goal of the CTBTO. Seismic monitoring of underground nuclear explosions (UNE) critically depends on the time-varying detection threshold (Kvaerna *et al*., 2007) associated with the primary seismic network of the International Monitoring System. The overall network resolution or individual sensitivity of IMS station(s) is directly transformed into the completeness of the Reviewed Event Bulletin. When monitoring UNEs at a global level, the (P)TS faces various regional types of natural seismicity, which present a challenge to the uniformity of detection and thus monitoring threshold. Higher seismic activity tends to make more difficult the task of smaller events detection.

Waveform cross correlation is one of the methods to enhance detection of the smallest events (Geller and Mueller, 1980; Israelsson, 1990; Joswig and Schulte-Theis, 1993). At regional distances, there are many studies showing significant improvement in detection (*e.g.* Schaff and Richards, 2005, 2011; Gibbons and Ringdal, 2006; Schaff, 2008; Schaff and Waldhauser, 2010) and location (*e.g*., Richards *et al*., 2006; Schaff *et al*. 2004; Schaff and Waldhauser, 2005; Waldhauser and Schaff, 2008; Yao *et al*., 2012) of smaller earthquakes. Withers *et al*. (1999) developed an automated system of detection and location of local and regional events based on cross correlation of envelopes. Gibbons and Ringdal (2004) applied cross correlation technique to multichannel signals generated by small, cavity-decoupled, underground tests and measured by array stations NORES, NORSAR, and Hagfors. Seven from eight explosions were too small to be detected by standard beamforming technique. Cross correlation allowed detecting all eight events. Following the first success of array based cross correlation, Gibbons *et al*. (2007) demonstrated substantial improvements in detection of signals and location of low-magnitude earthquakes at regional distances using multichannel waveform correlation at arrays station NOA. Harris and Dodge (2011) and Slinkard *et al*. (2013) applied array-based multichannel cross correlation algorithms to automatic recovery of aftershock sequences within relatively small footprints of intermediate size earthquakes.

There are two principal and linked features characterizing all regional studies using cross correlation. The length of template signal is usually larger than 20 s in order to include all regular regional phases, and the distance of cross correlation is limited to a few kilometers. The lengthy templates guaranty higher spatial resolution (i.e. relative location accuracy) and lower detection threshold at the expense of spatial coverage. Generally, a template of 50 s has the cross correlation distance less than 5 kilometers. This range is defined by the extraordinary high variability of regional wavefield, which includes not only the change in amplitude and phase spectra, i.e. the change in signal shape, but also increasing travel time differences between regular phases Pg, Pb, Pn, Sn, Lg as well as randomly changing coda. One would need millions of templates to cover the globe with regional master events. This is hardly a feasible task.

For teleseismic ranges, signals of interest are short (*e.g.* for explosions - just few cycles) and cross correlation is possible for sources spaced by 100 and more kilometers (Bobrov *et al*., 2012b; Kitov *et al*., 2012). This reduces the total number of master events in the global set to 25,000 (Kitov *et al*. 2013c). The problem of master events in aseismic areas can be resolved by the use of grand masters replicated over thousands of kilometers around their actual positions (Kitov *et al*., 2012) or synthetic templates (Rozhkov *et al.,* 2013). In this study, we assess the level of similarity between waveforms generated by events on the opposite ends of the Mid-Atlantic Ridge in order to justify the use of grand masters.

Specifically for the nuclear monitoring purposes, several studies have demonstrated a significant improvement in detection, location, and magnitude estimation of the announced

DPRK nuclear tests using waveform cross correlation when applied to teleseismic (Selby, 2010; Bobrov *et al.,* 2012) and regional/teleseismic (Gibbons and Ringdal, 2012; Schaff *et al.*, 2012) waveforms. Hence, global seismology and seismic monitoring may both benefit when waveform cross correlation is used by the IDC.

The completeness of the Reviewed Event Bulletin is a key quantitative characteristic of seismic monitoring under the CTBT. Currently, the REB is built in a pipeline consisting of automatic and interactive processing. The former creates a standard event list (SEL) by associating automatic detections using the full set of *Global Association* (GA) applications (Coyne *et al.*, 2012). To build the REB, analysts use information from various sources: the event hypotheses from the final automatic bulletin, which is called standard event list 3 (SEL3); appropriate automatic detections, not associated with the SEL3 events; and manually detected arrivals, which were missed in automatic detection. The arrivals associated with an REB event have to meet a number of quantitative requirements defining the quality of the event. In essence, a qualified REB event has to contain at least three primary IMS stations with primary phases (*e.g.*, Pg, Pn, and P) having the estimates of arrival time and vector slowness within the predefined (phase and station-dependent) uncertainty limits (Coyne *et al.*, 2012).

Here, we assess the performance of waveform cross correlation in signal detection and event building at teleseismic distances. When recovering seismic process at a global level, the cross correlation technique should be based on a dense 2D grid of master events uniformly covering the earth with an approximately constant spacing of ~1°. Within the Atlantic Ocean, seismic activity is mainly concentrated along the Mid-Atlantic Ridge. Thus, it can be represented by a quasi-1D distribution isolated from other seismic regions. This facilitates the process of master event selection, data processing, and interpretation of results. We expect that the catalog of earthquakes for the Atlantic Ocean obtained using cross correlation to be a complete one to the extent the full set of IMS seismic data allows. The events at the opposite ends of the Mid-Atlantic Ridge are separated by approximately 13,000 km. Here, one can explore the level of similarity between waveforms generated by physically similar sources located in practically antipodal points.

We have already analyzed a relatively short aftershock sequence in North Atlantic with the main shock on October 5, 2011 (Bobrov *et al*., 2012a). This earthquake had magnitude mb=4.23. The sequence contained 38 events in the REB. Also, there were three small foreshocks approximately one hour before the main shock. In interactive processing, an experienced analyst added 26 (68%) new events to the REB using cross correlation detections which had not been found in routine automatic processing. Therefore, the REB for this aftershock sequence was not a good bulletin and misses at least 70% of valid events. In turn, these valid events are missing from the ISC and not available for the researchers. This is also a bigger challenge to seismic monitoring under the CTBT.

## Data and method

The Atlantic Ocean is characterized by a relatively low seismicity despite active tectonics of the Mid-Atlantic Ridge and associated transform faults. The International Data Centre reports a few hundred events per year. With the growing number of primary stations the sensitivity of the IMS has been improving since 2001. Table 1 shows that the International Seismological Centre (2013) reports more events by a factor of 1.3 (2001) to 3.6 (2005). The years of 2005 and 2006 look like outliers with the number of events reported by the ISC above 1,000. Since 2007, this factor has been varying between 1.4 and 1.8. The total number of earthquakes in 2011 and 2012 is 1152 (576+576). The IDC catalogue for seismic region 32 is complete only to magnitude 4.3 to 4.5 (Kitov *et al*., 2011). The reason for this high detection threshold for the Atlantic Ocean consists in the absence of regional stations and large amplitude of the ambient microseismic noise generated within this region.

To characterize the level of correlation between signals from earthquakes within the Atlantic Ocean we have selected all REB events having arrivals at three and more primary IMS array stations in 2011 and 2012. In total, there were 931 such events or 81 per cent of the total number. Figure 1 presents geographical distribution of the selected events. This Figure also displays the configuration of twenty two involved primary array stations. Stations NOA and ESCD are excluded from our analysis because of problems with data quality and processing (Bobrov *et al*., 2012b). There are several auxiliary IMS arrays (*e.g*., EKA, HFS, SPITS, BVAR, KURK) detecting P-waves from the Atlantic events but not included because their data are not continuous. Twenty two stations reported 6,362 arrivals from 931 events. Figure 2 displays the frequency distribution of the number of primary IMS arrays, NSTA, and that of (IDC) body wave magnitude for the selected 931 REB events. One third of events have NSTA=3 and 4. Figure 3 depicts the aggregate frequency distributions of SNR for all 22 stations and three distributions for stations with the largest number of detections. Station TORD has 869 arrivals, AKASG – 632, and TXAR – 551 arrivals. It is worth noting that the curve for TORD has the slope lower than the other two stations. This observation may reflect better noise suppression due to larger aperture, signals of higher amplitude from the same sources, and more efficient beam forming at TORD.

Table 1. The number of events reported by the ISC and IDC for Flinn-Engdahl seismic region 32 (Atlantic Ocean).

| Year | ISC | IDC | IDC/ISC |
|------|-----|-----|---------|
| **2012** |     | 576 |         |
| **2011** |     | 576 |         |
| **2010** | 943 | 537 | 0.57    |
| **2009** | 699 | 501 | 0.72    |
| **2008** | 664 | 423 | 0.64    |
| **2007** | 936  | 590 | 0.63 |
| **2006** | 1146 | 440 | 0.38 |
| **2005** | 1137 | 317 | 0.28 |
| **2004** | 733  | 454 | 0.62 |
| **2003** | 563  | 412 | 0.73 |
| **2002** | 533  | 393 | 0.74 |
| **2001** | 541  | 410 | 0.76 |

For cross correlation detection, we selected sixty master events in the North Atlantic. All these master events contain high-quality waveform templates (signals with SNR>5) at ten best stations with optimal azimuthal gap. We cross correlated these templates with continuous waveforms in two half-year-long intervals in the beginning of 2009 and 2012. All waveforms were checked for quality problems like spikes, zeroes, missed intervals and/or channels, high noise levels and so on by standard IDC methods (Coyne *et al.,* 2012). Damaged records were recovered where possible. Problem channels or unrecoverable data intervals were excluded from processing.

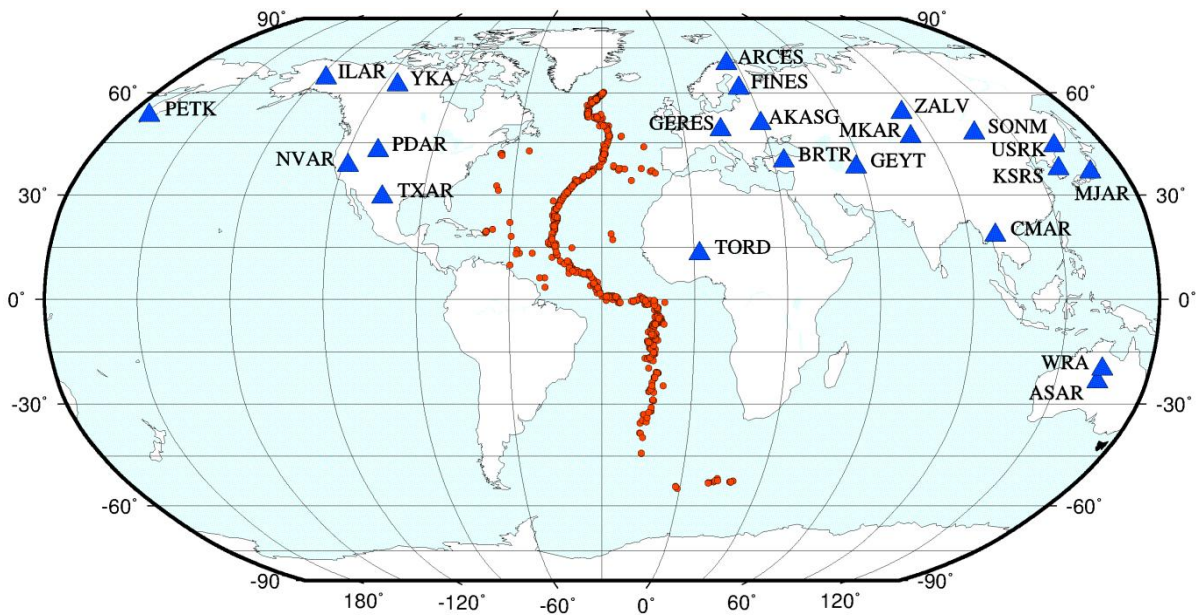

Figure 1. Locations of 931 REB events with three and more IMS array stations in Flinn-Engdahl seismic region 32 (Atlantic Ocean) in 2011 and 2012. There are 685 events in the northern hemisphere (74%).

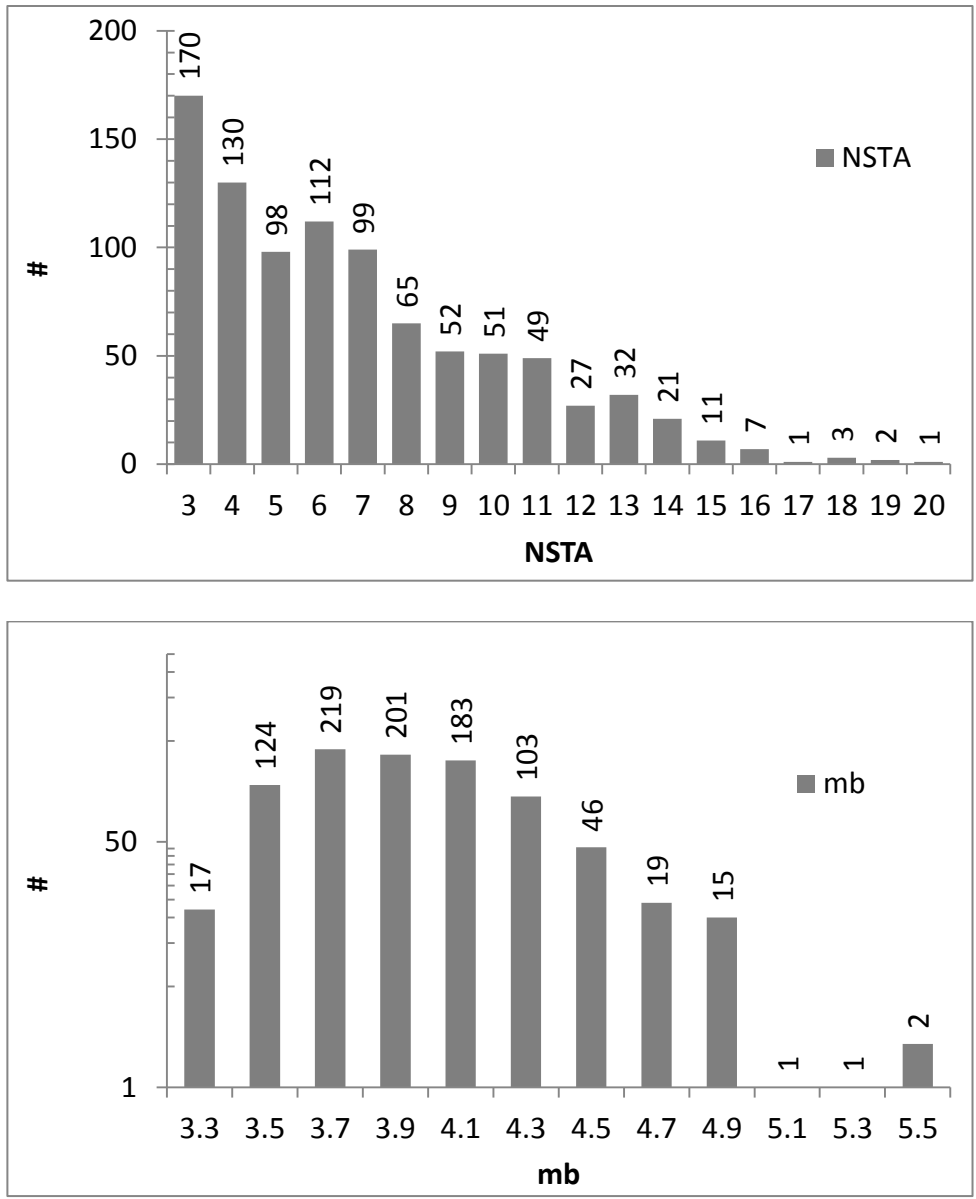

Figure 2. Frequency distribution of the number of primary IMS arrays, NSTA, and body wave magnitude for 931 REB events. Only events with NSTA>3 were selected from the REB.

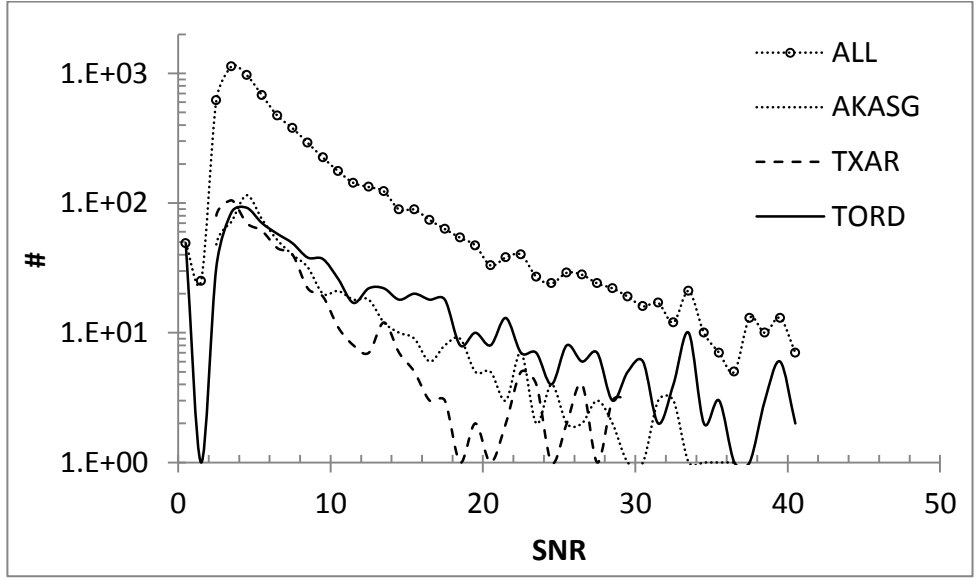

Figure 3. Frequency distribution of SNR for all arrivals in 931 REB events and three frequency distributions for AKASG, TORD, and TXAR.

To calculate a single time series of cross correlation coefficient, *CC*, between a multichannel template and continuous waveform at a given station, we estimated CC time series at individual channels and then averaged them as proposed by Gibbons and Ringdal (2006). This procedure allows smooth exclusion of poor channels without loss of continuity (Bobrov *et al*., 2012). Teleseismic signals from low-magnitude events are generally short and hardly include any phase except the primary one. Therefore, we use waveform templates containing only several seconds of the P-wave and a short time interval before the signal. To enhance detection of weak signals all templates and continuous waveforms are filtered in four frequency bands, and cross correlation coefficients are calculated in parallel. For the low-frequency (BP, order 3) filter between 0.8 Hz and 2.0 Hz, the total template length is 6.5 s and includes 1 s before the arrival time. For the intermediate frequency band 1 Hz to 3 Hz, the length is 5.5 s including 1 s lead. For two high-frequency filters 2 Hz to 4 Hz and 3 Hz and 6 Hz, the full length is only 4.5 s. In other words, the template should include 4 to 6 signal periods in a given frequency band. Taking into account that the number of vertical channels varies from 6 (BRTR) to 24 (WRA) one has from ~30 s to ~160 s long templates to cross correlate with the segments of continuous waveform of the same length. The time shifts between arrivals at individual channels provide very high sensitivity of the template records to azimuth and slowness of correlating signals.

The aggregate *CC*-traces are used to detect signals with standard short-term/long-term moving average ratio (STA/LTA). This ratio is also considered as signal-to-noise ratio characterizing *CC* signals, *SNR_CC*. The STA and LTA length is 0.8 s and 40 s, respectively. The detection threshold is defined by *SNR_CC*>2.5, but no detection with |*CC*| <0.2 is possible. Both thresholds have been estimated in our previous studies (*e.g.* Bobrov *et al.*, 2012b) in order to produce high quality cross correlation detections while retaining the rate of missed signals at a low level.

Cross correlation is a nonlinear transformation, which is also not a bijective function. Therefore, *CC*-traces at individual channels have no one-to-one correspondence with the original waveforms. However, having a set of individual *CC*-traces one can formally apply standard array processing and obtain meaningful results. Moreover, there are some advantages associated with CC array processing. All CC-traces are always in ±1.0 range and they do contain the portion of signal coherent with that of the template, i.e. the incoherent noise and signals are effectively suppressed. These conditions are favourable for *f-k* analysis, which is practically confined to a small footprint around the azimuth and slowness of the template signal. Gibbons and Ringdal (2006, 2012) proposed and successfully applied *f-k* analysis to the cross correlation time series at array stations. This afforded efficient rejection of signals with different vector slowness. For this reason, pseudo-azimuth and pseudo-slowness is estimated in automatic cross correlation processing using *f-k* for correlation time series for all detections. The term "pseudo" expresses the absence of one-to-one correspondence between the ground motion and CC domains. When the deviation from the master's azimuth and slowness is above some predefined thresholds the detection under investigation is rejected.

Traditional body wave magnitude cannot be estimated for a detection not associated with an event because of the absence of origin information. For a signal detected by cross correlation using a master template at a given station, Gibbons and Ringdal (2006) introduced an amplitude scaling factor,

$$\alpha = s \cdot m / m \cdot m, \qquad (1)$$

where $s$ and $m$ are the vectors of data for the slave and master event, respectively. This factor takes into account only the portion of the sought signal correlating with a template. For stations at regional distances, Schaff and Richards (2011) demonstrated that $\alpha$ is a more accurate and reliable measure of the relative size of a slave event compared to the master event when they are close. For teleseismic distances, the spacing between well correlating slave and master events may reach tens of kilometres and their signals may have different shapes. Then the assumption behind (1) is not valid and the relevant estimates are biased. Bobrov *et al*. (2012b) proposed to use $|s|/|m|$ instead of $\alpha$. The logarithm of this ratio,

$$RM = \log(|s|/|m|) \qquad (2)$$

is the magnitude difference or the relative magnitude. This definition is also subject to bias when the slave signal is close to noise in amplitude. For SNR < 2.0, equation (1) may effectively replace (2) since cross correlation can detect only weak signals similar in shape from the slaves close to the master in space.

In a multichannel template waveform, the absolute arrival times of the master signal at individual sensors of a given array station are delayed relative to the arrival time at the reference sensor as defined by their positions and by the master/station slowness and azimuth

$$dt_i = \boldsymbol{w} \cdot \boldsymbol{d_i} \tag{3}$$

where $\boldsymbol{w}$ is the vector slowness of the P-wave arrival at the reference array sensor, and $\boldsymbol{d_i}$ is the vector between the reference and *i*-th sensor. The change in $\boldsymbol{w}$ over the array is neglected in (3). The time delay can be positive or negative. For the purposes of cross correlation processing, we shift the individual traces to the reference channel according to their theoretical delays, $-dt_i$, calculated for a plane wave with the master/station vector slowness. Then these absolute arrival times are synchronized to the extent the theoretical and empirical time delays coincide. This operation is needed to provide a uniform presence of the template signal at all channels. It is usually neglected in regional cross correlation studies with template windows of tens of seconds. At the most remote sensors, the absolute arrival times for a teleseismic P-wave with slowness of 0.07 s/km may lag or lead the reference arrival time by more than one second. For a 4.5 s template window, one second of missed signal is crucial. In the continuous waveforms, the same shifts are applied in order to synchronize them with the template. This is a standard setting for detection since the sought slave events are assumed to have practically the same time delays as in the template.

At array stations, actual arrival times at individual channels may deviate from the theoretical ones. These deviations are fully related to propagation of plane waves in the inhomogeneous velocity structures beneath the arrays. They reduce the effectiveness of beam forming and result in a significant beam loss (Coyne *et al*., 2012). Due to spatial proximity, the slave events should have practically the same azimuth and scalar slowness as the master event at a given station. Therefore, cross correlation between a master and continuous waveform is not subject to the effects similar to beam loss. Cross correlation should be superior in signal enhancement and detection.

Having the set of detections obtained using cross correlation for a given master we can associate these arrivals and build events. The Global Association currently used at the IDC is based on a sophisticated set of algorithms (Coyne *et al*., 2012). The GA plays with an enormous number of hypotheses, which may link one arrival with various events worldwide. The association of cross-correlation detections obtained by one master is much simpler. We are looking for slave events in a small footprint of the master. Thus, all qualified phases are associated with events local to the master. This process is called "local association", *LA*.

For *CC*-detections, there are estimated arrival times at each station, $at_{ij}$, where $i$ is the index of the $i$-th arrival at station $j$. Under the *LA* framework, it is assumed that all valid arrivals have to be generated by slave events near the master one. For a master in a fixed location, one can estimate theoretical travel times, $tt_j$, to the involved stations. It is instructive to use *ak135* velocity model, which is also used in the current version of IDC processing. Apparently, the travel times from the sought slave events to the relevant stations can be accurately approximated by the master/station travel times. Using these approximated travel times (same for all events around the master) and the measured arrival times one can calculate origin times for all detections:

$$ot_{ji} = at_{ij} - tt_j \qquad (4)$$

For the master, one has a set of origin times at the stations instead of the arrival times. To match the full set of event definition criteria (EDC) related to the REB (Coyne *et al*., 2012) one needs three or more origin times at different stations to group within a few seconds.

Since the REB requires travel time, azimuth, and slowness residuals as well as station/network magnitude differences within the phase and station related uncertainty bounds we process the original waveforms and estimate the whole multitude of standard parameters for the cross correlation detections. This information is also used by analysts during interactive review.

The suggested global grid of master events is sparse (~1° spacing) for the purpose of accurate location. To reduce the influence of master/slave distance on the scattering of the origin times, two concentric rings of 6 and 12 virtual masters at distances 0.225° and 0.450° from each and every master are introduced. The theoretical master/station travel times are recalculated for these eighteen virtual masters and nineteen sets of origin times are created for one master. These sets are used to build slave hypotheses by the local association algorithm. Hence, one may have from 1 to 19 event hypotheses based on the same set of CC-detections. We select the one with the largest number of associated stations and the lowermost origin time RMS residual. There can be specific conflicts between defining parameters ($RM$, azimuth, slowness) of the arrivals in the same hypothesis. When they are resolved, a set of REB-consistent hypotheses for the master event is obtained.

In the global grid, many adjacent masters may find the same event using cross correlation. To resolve this conflict, we also chose the event with the largest number of stations and the smallest origin time RMS residual from the set of conflicting hypotheses. The arrivals associated with rejected hypothesis are removed from the list of detections, which can be used iteratively. With all conflicts resolved, we obtain the final XSEL version, which is the start point of interactive processing.

At this stage, the XSEL events should just formally meet the EDC, and thus, interactive review of these events is mandatory. However, the XSEL events are characterized by quantitative parameters related to cross correlation, which can be interpreted in terms of event quality or the probability of hypotheses. This allows formulating a probabilistic approach to XSEL hypotheses similar to that underlying the EDC. For example, the cross correlation coefficient averaged over all detections associated with a given event, *CC_AV*, characterizes the average quality of these arrivals and the event as a whole. The cumulative cross correlation coefficient, *CC_CU*, can be a proxy to the overall event quality. Together with the number of associated stations and various individual characteristics of associated arrivals we use *CC_AVE* and *CC_CU* in a machine learning exercise. To create a training set of valid arrivals, we use those XSEL events, which are matched by origin times in ISC catalogue. Invalid arrivals are randomly chosen from the XSEL events not matched by the ISC. When the learning step is complete, we apply the estimated classification model to the XSEL.

## Cross correlation of REB events

At first, we assess the level of cross correlation between signals generated by earthquakes in seismic region 32 and detected by the IDC. Each and every REB event has detections obtained by standard IDC tools and reviewed by analysts. This procedure practically guarantees the presence of clear signals with appropriate travel time, azimuths, and slownesses residuals. Generally, the P-wave arrivals are characterized by SNR above 2.0. For the sake of quality we removed all SNR<2 arrivals from the following analysis.

Since we do not need to re-detect already existing REB arrivals cross correlation between two events at a given station is a straightforward procedure. In a given pair, one event is first considered as a master and then as a slave. The master template is defined by the relevant arrival time in the REB. Cross correlation coefficient is calculated using the slave continuous waveform with individual channels synchronized according to the theoretical azimuth and slowness of the P-wave propagating from the master event to the station.

Therefore, the swap of master and slave may result in a different delays pattern. In this case, the estimates of CC are not master/slave symmetric. The calculation of *CC* starts 3 s before the slave arrival time in the REB and ends 3 s after it. When there is no REB arrival in the slave, we process the same 6 s interval around the relevant theoretical arrival time. In other words, when a master has REB arrivals at 15 IMS array stations and a slave only at 5, we estimate 15 CCs. When the master and the slave are swapped, only 5 CCs are estimated. All calculations are carried out in four frequency bands with the relevant template windows. The peak |*CC*| value characterizes the master/slave pair.

For a given array station, there are two different time delay settings to calculate *CC*, *SNR_CC*, and *RM* for two REB events. In a standard setting, we shift individual channels in the slave waveform according to the time delays in the master template. We call this setting "master/master" (MM) since individual traces in the master and slave are shifted according to the master time delays. This is the setting for detection using cross correlation.

When cross correlating signals from REB events in the Atlantic Ocean, one has locations separated by 120º. This suggests a different setting with individual channels of the slave shifted according to the slave/station azimuth and slowness. This setting is called "master/slave" (MS). In this setting, all channels in the slave waveform are always synchronized with the reference sensor. For the MS setting, cross correlation coefficient for a master/slave pair spaced by a few thousand kilometres has to be larger than in the MM setting.

Table 2 presents general arrival statistics for 931 REB events in seismic region 32 as reported by the IDC in 2011 and 2012. Column 2 lists the number of arrivals in the REB for 22 IMS array stations. As mentioned above, we selected only those events, which include at least 3 primary array stations with P-wave signals having SNR>2. This somehow guarantees signal quality to distinguish between the absence of similarity and noise influence.

Station TORD is the most sensitive to earthquakes in the Atlantic Ocean with 869 arrivals (93%) and the poorest station is USRK with only 7 arrivals (<1%). There are excellent array stations WRA and ASAR with poor statistics because the epicentral distance is beyond 100º for most of events in seismic region 32. Stations MJAR, PETK, USRK generally have poor detection statistics. Due to specific features of the ambient microseismic noise, station ARCES generally has the peak SNR at higher frequencies, and thus, is not sensitive to low-frequency signals generated by events in the Atlantic Ocean.

There are seven stations with more than 50% presence in 931 events highlighted bold in Table 2: AKASG, BRTR, GERES, ILAR, PDAR, TORD, and TXAR. Stations FINES, MKAR, and YKA (italicized in Table 2) demonstrate a relatively good performance and complete the list of ten stations used for building XSEL in the next Section. These ten stations can provide a good azimuthal coverage needed for the creation of robust event hypotheses. Three stations in the same azimuth would likely produce many invalid events and a special constraint on the maximum azimuthal gap is needed to suppress these hypotheses.

There are 865,830 possible pairs of events (we exclude 931 autocorrelation cases) and cross correlation coefficients can be organized as a (nonsymmetrical) 931×931 matrix. The number of stations with signals qualified for the use as a template waveform varies with event. Many of the studied REB events have only 3 stations with relatively weak signals. In total, there are 355,382 (45%) master/slave pairs with at least one station with correlating signals ($|CC|>0.2$) in the MM setting and 690,337 (80%) pairs in the MS configuration. Figure 4 displays the frequency distribution of the number of stations, nsta, with $|CC|>0.2$ for all possible pairs of events. Most of event pairs have 1 or 2 stations with correlating signals, with at least 3 templates in each pair.

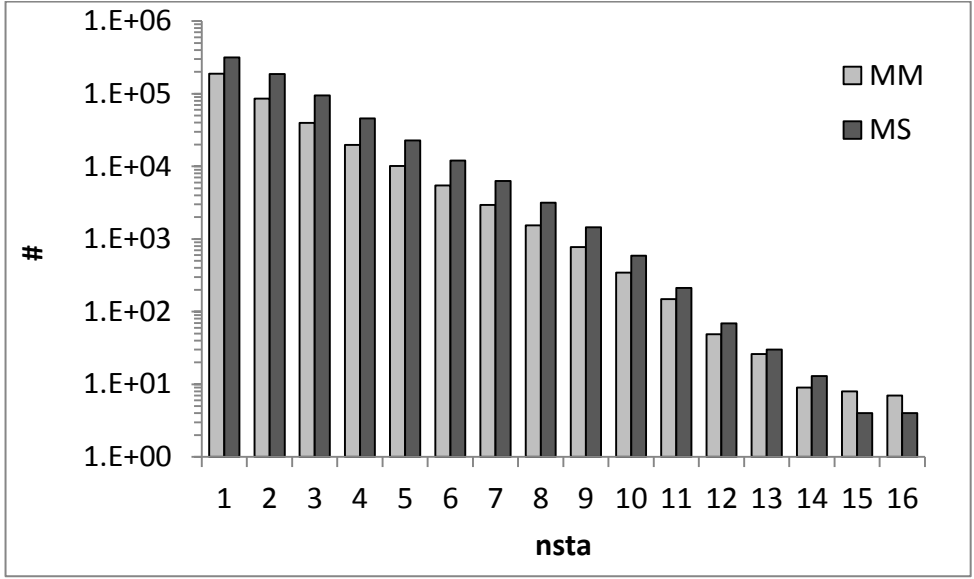

Figure 4. Frequency distribution of the number of stations, nsta, with |CC|>0.2 for all possible pairs of events. Notice the lin-log scale.

Table 2 also lists station-by-station statistics of cross correlation between all pairwise permutations of 931 REB events. Station TORD gives 144,009 absolute *CC*s above 0.2 from 755,161 (869×869) estimated (19%) in the MM configuration and 511,110 *CC*s (68%) above the threshold in the MS setting. There are a few possible reasons behind the relatively low rate of successful correlations. Firstly, most of signals are weak with smaller SNR and the ambient incoherent noise destroys correlation. Secondly, for the MS setting the deviations from the theoretical time delays may increase with the master/slave distance that ruins the coherency of *CC*s at individual channels and suppresses the aggregate *CC* estimate. Thirdly, for the MM setting, the increase in the master/slave distance leads to desynchronization of individual channels in the slave waveform where all channels are shifted by the master theoretical time delays. Fourthly, focal mechanisms and source functions change even between close events and may differ significantly for remote events. Fifthly, the change in geometrical spreading and nonlinear attenuation along varying propagation paths results in the change of signal shape. The latter two factors are crucial importance for the creation of the global grid of master events.

As expected, the MM setting is less effective for cross correlation of remote events. Stations TORD and AKASG demonstrate an increase in the number of |CC|>0.2 by a factor of 3.6. These two stations have relatively large apertures and the numbers of individual vertical sensors: 15 and 23, respectively. The other eight stations are less sensitive to the change from the MM to MS setting but also improve the overall statistics.

Table 2. Statistics for REB arrivals with SNR>2 (ARIDS) at 22 IMS stations. Cross correlation statistics in two configurations – MM and MS.

| STA | ARIDS | ARIDS/931 | MM | MS | MS/MM |
|---|---|---|---|---|---|
| **AKASG** | 632 | 0.68 | 55018 | 201524 | 3.66 |
| ARCES | 243 | 0.26 | 14464 | 22101 | 1.53 |
| ASAR | 18 | 0.02 | 165 | 162 | 0.98 |
| **BRTR** | 679 | 0.73 | 71629 | 133909 | 1.87 |
| CMAR | 28 | 0.03 | 420 | 369 | 0.88 |
| *FINES* | 415 | 0.45 | 14030 | 19020 | 1.36 |
| **GERES** | 481 | 0.51 | 35360 | 58641 | 1.66 |
| GEYT | 135 | 0.14 | 2315 | 2657 | 1.15 |
| ILAR | 498 | 0.53 | 86020 | 127472 | 1.48 |
| KSRS | 18 | 0.02 | 95 | 90 | 0.95 |
| MJAR | 10 | 0.01 | 45 | 46 | 1.02 |
| *MKAR* | 332 | 0.36 | 49696 | 55443 | 1.12 |
| NVAR | 190 | 0.20 | 4839 | 5965 | 1.23 |
| **PDAR** | 505 | 0.54 | 79908 | 87366 | 1.09 |
| PETK | 9 | 0.01 | 69 | 65 | 0.94 |
| SONM | 184 | 0.20 | 11966 | 11624 | 0.97 |
| **TORD** | 869 | 0.93 | 144009 | 511110 | 3.55 |
| **TXAR** | 551 | 0.59 | 75072 | 106139 | 1.41 |
| USRK | 7 | 0.01 | 37 | 28 | 0.76 |
| WRA | 18 | 0.02 | 38 | 41 | 1.08 |
| *YKA* | 450 | 0.48 | 38304 | 86362 | 2.25 |
| ZALV | 163 | 0.17 | 4300 | 4658 | 1.08 |

Figure 5 displays the distribution of distance between the REB events which are ordered by latitude from north to south. A few aftershock sequences and areas with repeated events are seen near the diagonal. The size of these zones is proportional to the number of events in the relevant series. Almost each of 931 events has tens of events within a few hundred kilometers. For this quasi-linear density, the choice of master events can be based on the number of correlations within, say, five hundred kilometers. The event with the highest portion of correlating REB events in a given segment is likely the best master for detection and event building. Overall, such masters should not be closer than 100 to 200 km from each other to cover different but slightly intersecting areas. The set of masters in the Atlantic Ocean could serve as a large-scale test grid on the way towards the global grid.

It is reasonable to expect that, *ceteris paribus*, signals from the closest events are characterized by higher cross correlation coefficients. In Figure 6, we depict the matrix of *CC_AV* estimates as an aggregate measure of correlation between two REB events. For the MM setting, there is no gap in correlation between adjacent events and the level of correlation falls with distance. The events in the North Atlantic do not correlate well with those in the southern hemisphere.

The MS setting effectively retains theoretical synchronization between individual channels independently on the master/slave distance. However, the empirical deviations from the theoretical time delays do change with the master/slave azimuth and slowness difference. Even with some small differences in these empirical deviations, the level of correlation between the northern and southern events should be practically the same as for the closest events, when their signals are similar. Right panel of Figure 6 demonstrates that *CC_AV* for the MS setting is high for all distances. Therefore, it is feasible to move a master event from the northernmost part of the Atlantic Ocean to the southernmost one (~12,000 km), to change time delays at individual channels correspondingly, and to use it as a master event for a different region. This observation strongly supports our suggestion to populate vast aseismic areas with replicas of grand master events (Kitov *et al.*, 2012). For the Atlantic Ocean, it is possible to use one or few best masters to build a master event grid covering the whole seismic region 32.

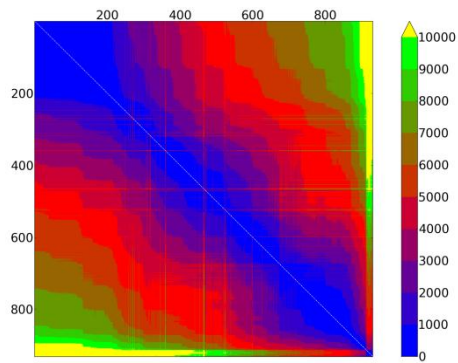

Figure 5. Distance between the REB events, which are ordered by latitude.

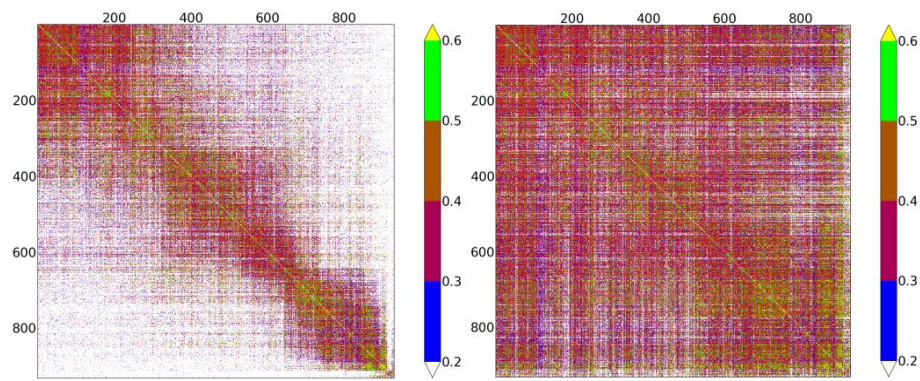

Figure 6. The average cross correlation coefficient, *CC_AV*, for 931×931 pairs of REB events. There are MM (left panel) and MS (right panel) configurations of time delays at individual channels. The REB events are ordered by latitude.

Figure 7 illustrates the quantitative gain of the MS setting for cross correlation of remote events. Having a limited effect at distances less than a few hundred kilometers, the MS setting improves the average *CC* by more than 0.3 for the events at distances of a few thousand kilometers. The cumulative *CC* generally gains more than 0.5, and for many event pairs more than 1.0. There are some event pairs with reduced estimates of *CC_AV* and *CC_CU*, mainly for closer events. This could be the result of poor piking of the involved REB arrivals and the resulting mislocation of smaller events. These smaller events might be much closer to the bigger ones than the IDC locates using standard tools. That makes the MS setting slightly biased and the MM presents a more accurate approximation.

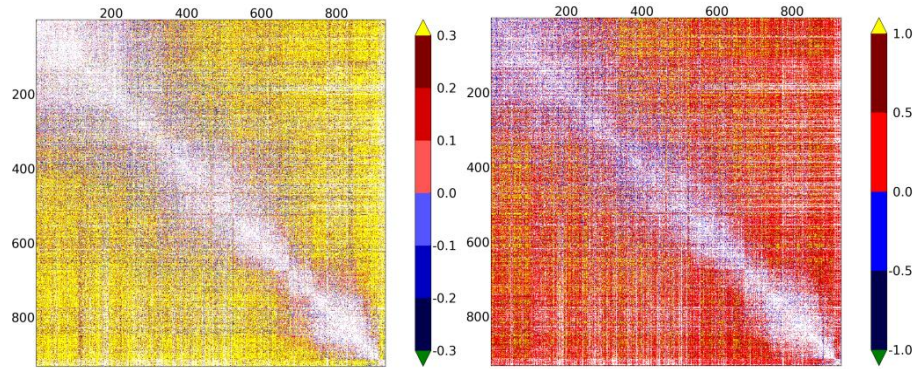

Figure 7. The difference between MS and MM matrices for *CC_AV* (left panel) and *CC_CU* (right panel). The zero differences are shown by white color.

Figure 8 displays |CC| matrices for six IMS stations. TORD is the best station with the largest number of correlating signals and relatively high *CC*s. Station BRTR is characterized by very high *CC*s but the number of correlations above 0.2 is lower than that for TORD. Overall, the distribution of *CC*s at stations AKASG is similar to that at BRTR, but the level of correlation is by 0.2 lower. TORD and BRTR cover the whole length of the Mid-Atlantic Ridge. On the contrary, all IMS stations within North America (ILAR, PDAR, TXAR, and YKA) do not provide CC estimates for the southern events because of the P-wave propagation shadow zone beyond ~100º. As a result, the coverage of the North Atlantic with IMS stations is much better than that for the most southern part of region 32. Stations AKASG and GERES provide uniform coverage except the southernmost segment of the Mid-Atlantic Ridge with a few events.

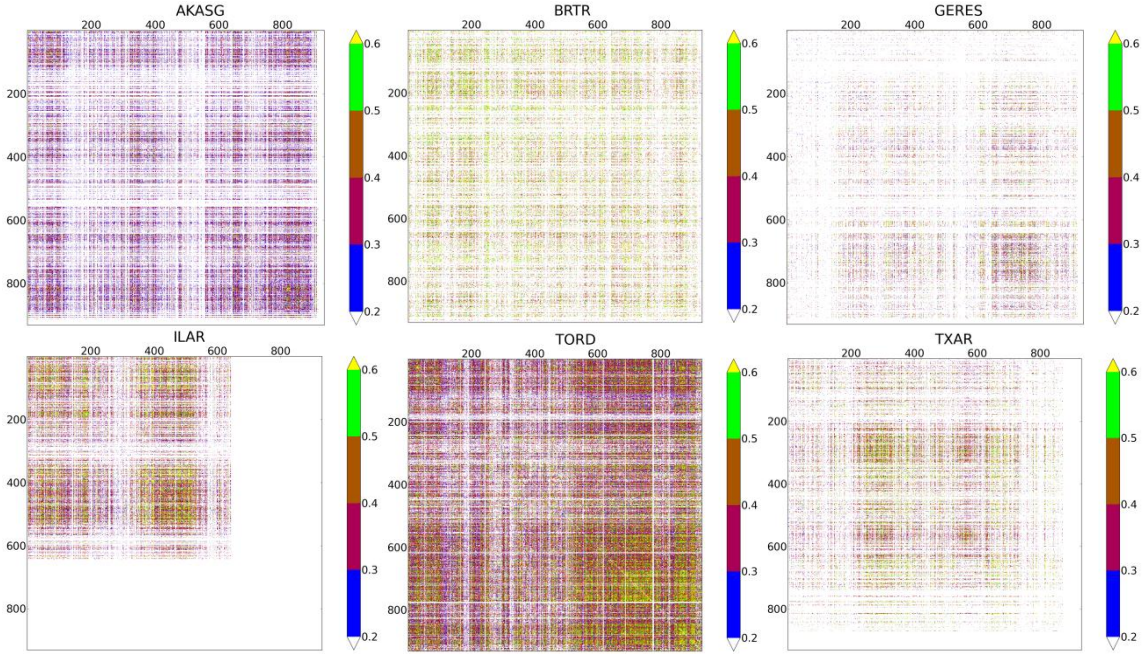

Figure 8. Absolute *CC* for the MS setting for six IMS stations with the highest number of correlating events.

Having estimated all cross correlation coefficients for all REB pairs one can calculate the frequency distribution or the probability density function (PDF) for the average and cumulative *CC* for the pairs where at least one station has |CC| > 0.2. Figure 9 illustrates this result and shows that, in relative terms, the MS setting is slightly more efficient between 0.3 and 0.5. For the cumulative *CC*, the MS setting gives relatively more events between 1.0 and 3.5.

Figure 10 presents two important characteristics of CC-detections: *SNR_CC* and *CC_STDEV*, which is the standard deviation of the *CC* estimates at individual channels. Two *SNR_CC* distributions reveal that the MS setting provides a higher portion of larger SNRs measured from the CC-traces. The superiority of detection using cross correlation is clear when SNR_CC and standard SNR are compared for the same arrival set. The SNR curve in Figure 10 estimated from 6,436 arrivals associated with 931 REB events is lower than both SNR_CC curves in the range between 2.5 (minimum SNR_CC) and 4.8. Approximately 10% of these REB arrivals are characterized by SNR<3 (28% have SNR<4 and 43% SNR<5). For these arrivals, the gain in SNR provided by cross correlation is most prominent. By definition, |CC|≤1.0 and SNR_CC cannot reach larger values since the RMS level of noise CC fluctuates around 0.1 for IMS stations. Therefore, standard SNR is larger than SNR_CC for high-amplitude signals.

The peak of the *CC_STDEV* curves is between 0.15 and 0.2 for both settings. This might be too large value for the 0.2 threshold and we are going to use *CC__STDEV* as a parameter for rejecting unreliable arrivals. Obviously, the reliability of *CC* estimates depends on *CC_STDEV* – the detections with low *CC*s and high *CC_STDEV* are likely false alarms. The events consisting of a few arrivals with unreliable CC-detections are more probable to be invalid. Figure 10 also demonstrates that the MS-related PDF for *CC_STDEV* has slightly higher amplitude above 0.3, but otherwise both distributions practically coincide. It is worth noting that 80% of CCs have *CC_STDEV* <0.2 and only 1% have *CC_STDEV*>0.3.

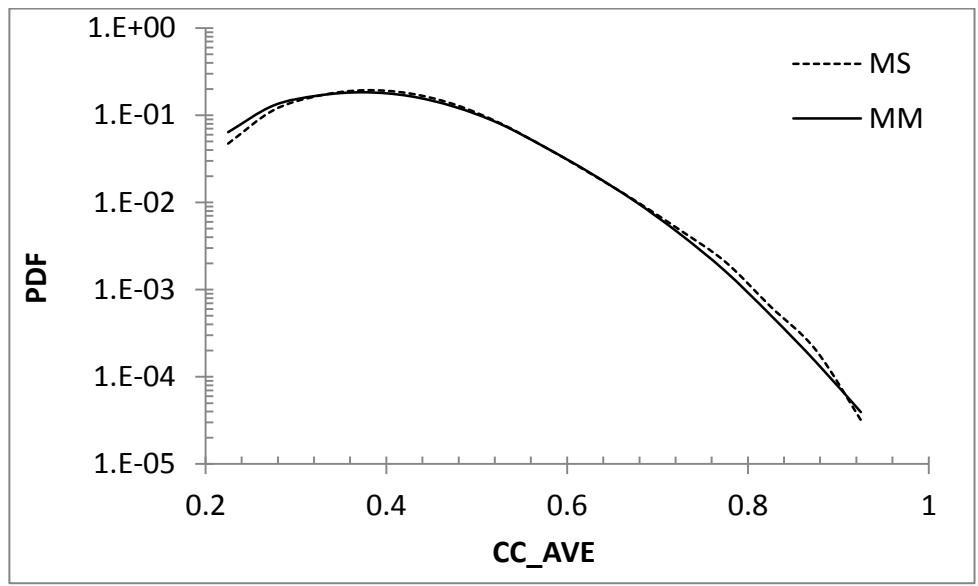

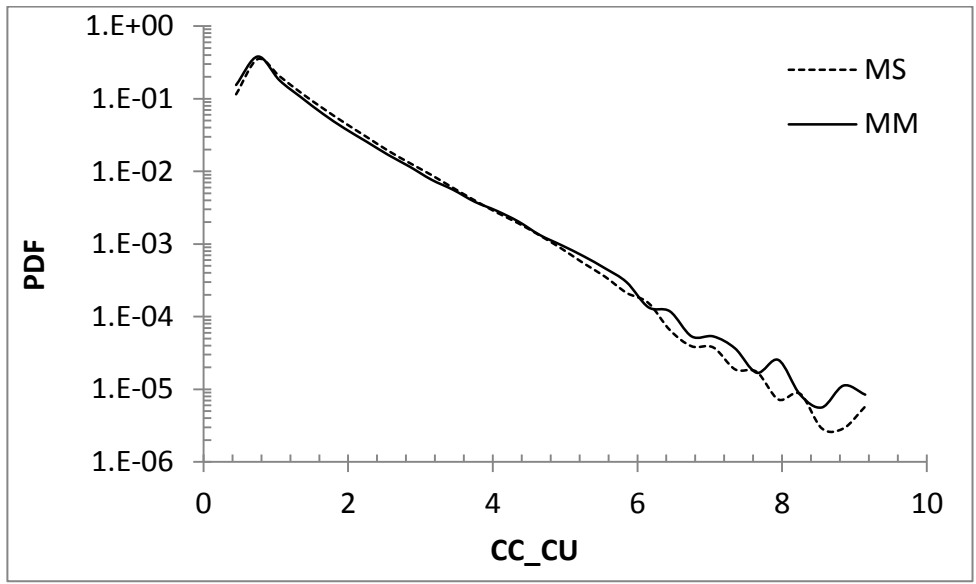

Figure 9. PDF of *CC_AV* and *CC_CU* for all qualified pairs

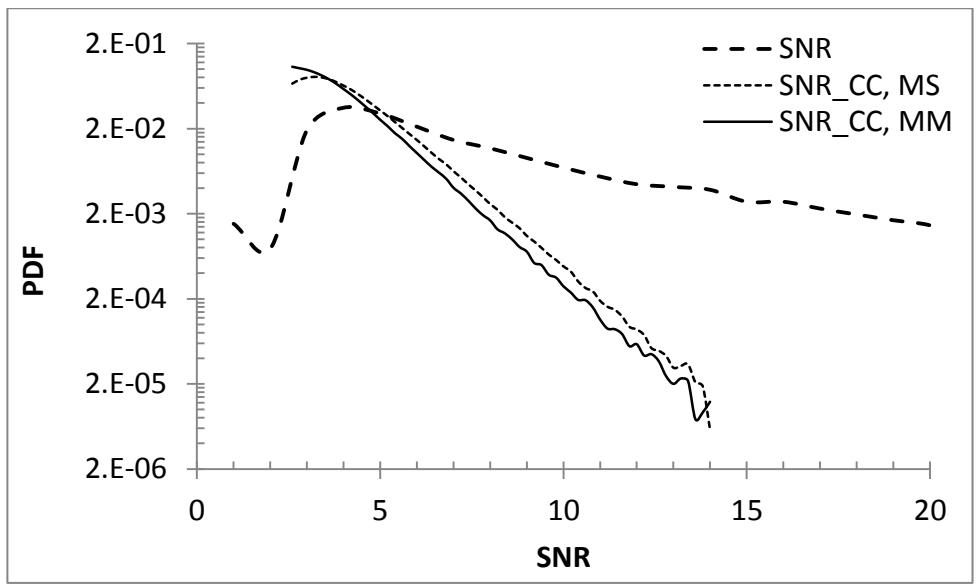

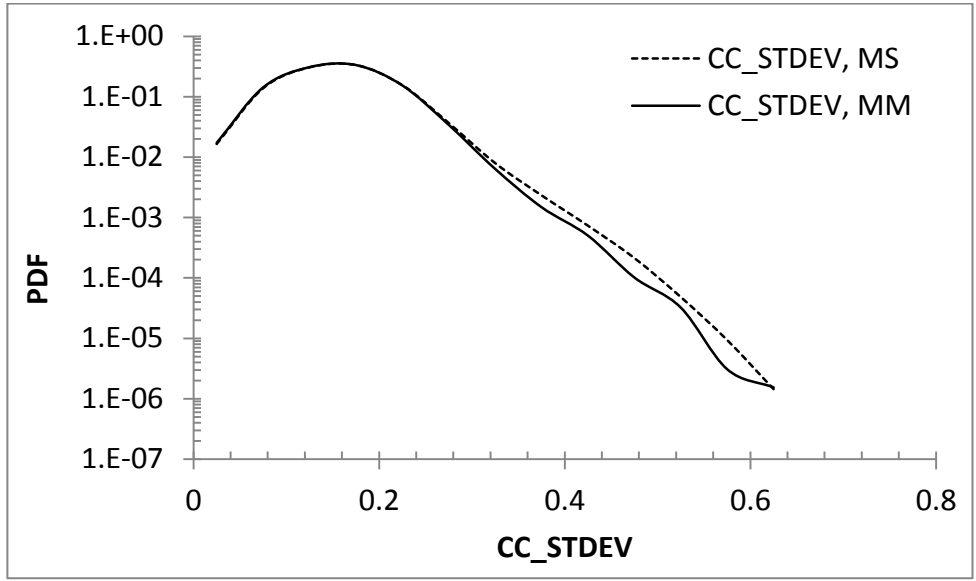

Figure 10. PDF of SNR vs. SNR_CC and CC_STDEV for all detections in all pairs of correlating REB events. Notice the lin-log scale.

Before we proceed to detection and event building in the North Atlantic it is worth describing the performance of the involved IMS stations. Figure 11 depicts ten PDFs for *CC* as estimated for 931 REB events. There are three stations with the PDFs different from other seven: GERES, YKA, and BRTR. The latter curve has a prominent peak at 0.45, which is absent in other curves. The other two stations show a much faster fall in the relevant PDFs, which is compensated by an increased density of positive (GERES) or negative (YKA) CCs.

There are four IMS stations in North America and six on the other side of the Atlantic Ocean. Overall, the azimuthal coverage is not perfect and is characterized by low sensitivity to epicenter movement along the north-south direction. Moreover, station triplets on one side of the Mid-Atlantic Ridge can produce many false hypotheses based on similar errors in master/station travel times. A similar effect is observed in standard IDC processing when three-station event hypotheses with poor azimuthal coverage (large azimuthal gap) have locations with confidence ellipses of 10,000 km$^2$ and aspect ratios of 1:10 (Pearce *et al.*, 2011). However, one should be careful when rejecting such triplet-based events. Many of them may be valid and likely not seen on the other side due to earthquake directivity.

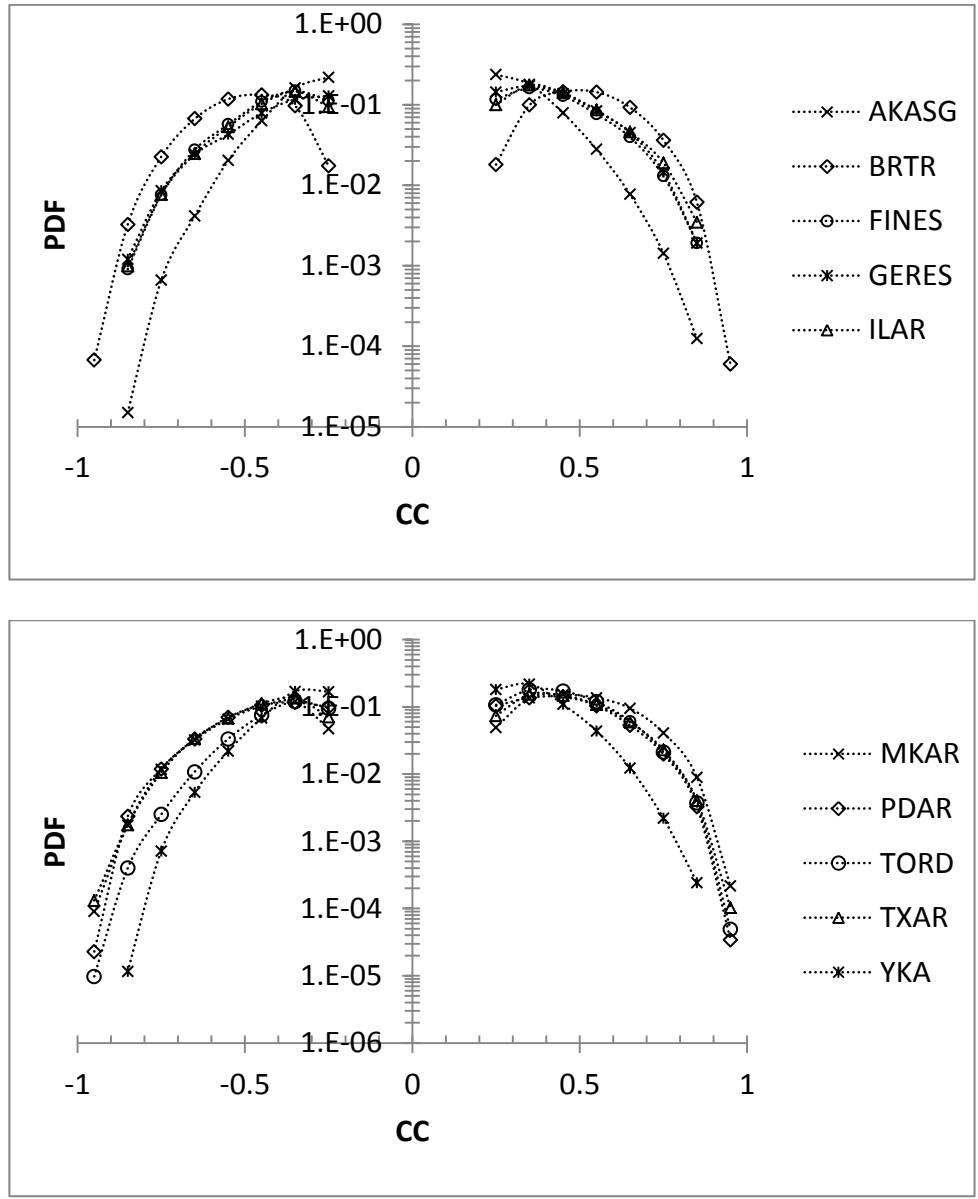

Figure 11. PDF for CC at ten IMS stations.

Summarizing the results of cross correlation of 931 REB events in the Atlantic Ocean we have to highlight several major features. Firstly, the current IDC detection tool is not very sensitive to earthquakes in Flinn-Engdahl seismic region 32 and the relevant part of the REB is likely complete to mb(IDC) 4.1 to 4.5. This value can be considered as a monitoring threshold (Kvaerna *et al*., 2007; Coyne *et al*., 2012). As in other studied seismic regions, waveform cross correlation based on real master events should be able to reduce the monitoring threshold by 0.3 to 0.4 units of magnitude (Bobrov *et al*., 2012ab; Kitov *et al*., 2012). Secondly, one can select a set of master events from the REB, which are tightly (~1 º) covering the Mid-Atlantic Ridge line and well correlating with neighboring events. Moreover, the events from the opposite side of the Atlantic Ocean might be also used as masters. This allows reproducing of the best masters over a 2D grid extended by thousands of kilometers from the Mid-Atlantic Ridge. Thirdly, there are several REB events poorly correlating with neighboring events. These events are likely mislocated or built using wrongly associated phases. Fourthly, we selected ten IMS array stations with good detection capability for the events in the studied region. However, they have quite different sensitivity. Therefore, the event hypotheses based on the best stations (*e.g*., AKASG, BRTR, TORD, and TXAR) are more likely to be valid than those containing only the stations with lower sensitivity (*e.g*., FINES, MKAR, and YKA).

## Finding events in the North Atlantic

In this Section, our interest is focused on the North Atlantic due to its higher seismicity as measured by the IDC. We have selected sixty master events from the REB and retrieved corresponding waveform templates at ten IMS array stations from the IDC archive database to cover the Mid-Atlantic Ridge between 3ºN and 60ºN. For simplicity, there are no master events off the ridge since seismic activity within other parts of the Atlantic Ocean is low or local. In order to assess the relative performance of master events we intentionally allowed them to have varying number of templates: from 5 to 10. All signals used as templates have SNR>5 to reduce noise influence on cross correlation. Figure 12 depicts the frequency distribution of body wave magnitudes for 60 masters as estimated according to the IDC definition. Six events have mb(IDC)<4.0. Due to similar frequency content, the waveforms from these lower magnitude masters may better correlate with the signals from the smallest events we are searching. Right panel of Figure 12 displays the number of templates with SNR>5 for 10 stations. The best one is ILAR with 59 templates from 60. This does not contradict Figure 8, where ILAR demonstrates a higher level of cross correlation for the North Atlantic events. TORD and BRTR have 57 templates each. MKAR and GERES are the poorest stations from these ten.

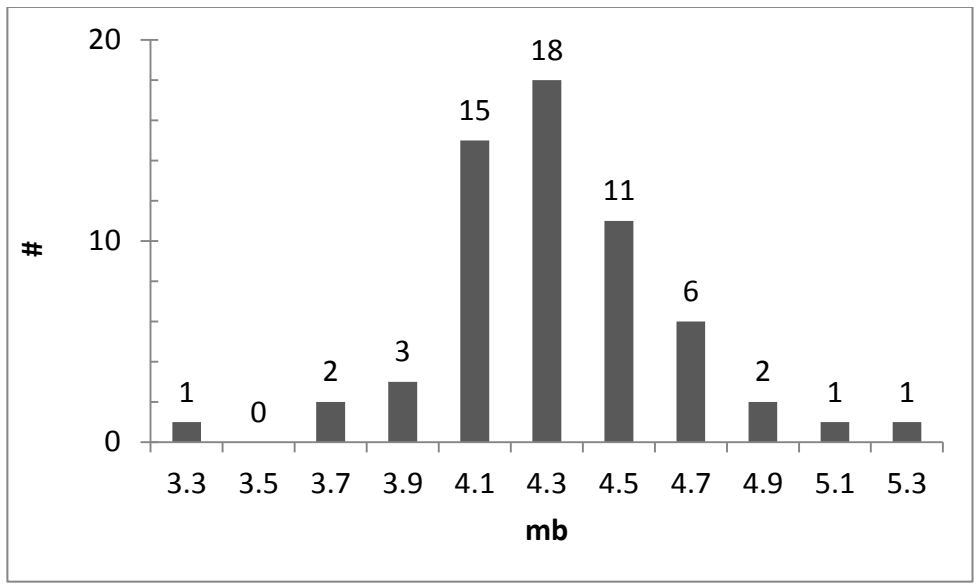

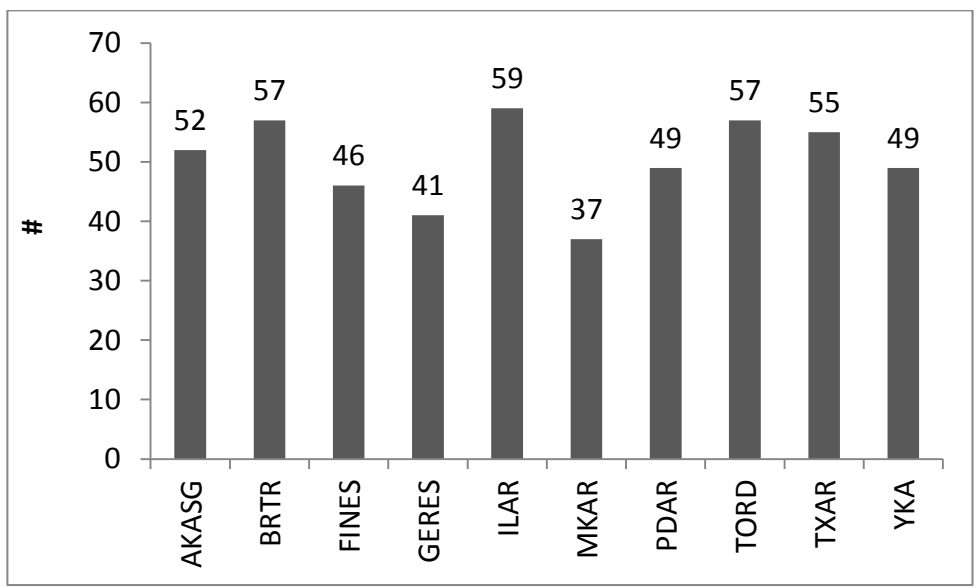

Figure 12. Frequency distribution of magnitudes in 60 master events (left panel) and the number of templates with SNR>5 in 60 masters for ten stations (right panel).

As in our previous studies (Bobrov *et al*., 2012ab; Kitov *et al*., 2012), we put the detection thresholds low: *SNR_CC*>2.5 and |*CC*|>0.2. Standard STA/LTA detector was applied to continuous cross correlation traces. The azimuth and slowness residuals for the detections associated with an XSEL event should not be out of ±20º and ±2 s/deg, respectively, from their theoretical values. The station/network *RM* residual should be less than 0.7. The *LA* algorithm includes a constraint on the maximum azimuthal gap of 270º for XSEL events.

At first, we processed the first half of 2009 and obtained 195,339 detections at ten stations. There were 64,063 event hypotheses built in the *LA* process. After the conflict resolution between adjacent masters an XSEL containing 58,443 events was built with 177,520 associated arrivals (see Table 3 for details). This XSEL includes 91 per cent of the total number of detected events and arrivals. Therefore, there is only a small portion of detections, which cannot be associated with events using the origin time window of 6 s for one master. This is a good indication that the involved IMS stations did detected valid events.

There were only 135 events within the North Atlantic reported in the REB during the same period. This enormous difference can be explained by various reasons. To exclude time specific features, we processed the first half of 2012 and built 46,804 XSEL events with 184 events in the REB for the same period. Therefore, the North Atlantic seismicity is likely higher than that reported by the REB, but the XSEL may contain many invalid events.

Table 3. Statistics of detections and XSEL events

|  | Detections | | | Events | | |
| --- | --- | --- | --- | --- | --- | --- |
|  | All | Assoc. | % | All | After CR | % |

| 2009 | 195339 | 177520 | 91 | 64063 | 58443 | 91 |
| 2012 | 178875 | 142167 | 79 | 58751 | 46804 | 80 |
| 2012_20 | 12899 | 8838 | 69 | 4048 | 2830 | 70 |

The Mid-Atlantic Ridge is one of tectonically active regions of the world. The seafloor spreading driven by mantle convection creates fresh crust and lithosphere in the ridge zone. This process is accompanied by extremely high heat flow and temperature gradients as well as active tectonic movements in the crust subject to brittle fracturing. This is likely a permanent process generating low-amplitude signals (or practically continuous and noise), which are measured by the involved IMS arrays steered to the masters events. Due to the permanent sources, this noise likely has an elevated coherency at teleseismic stations.

The reliability of a detection obtained by a given waveform template critically depends on the template length relative to the length of the slave signals. Routinely, we use frequency dependent time widows from 3.5 s to 6.5 s. For a middle-aperture ten-element array, these windows are enough to suppress random noise. For the coherent noise from the Mid-Atlantic Ridge, we increase the length of all widows approximately by a factor of four with the longest window of 20 s. This approach reduces the number of events built for the first half of 2012 down to 2,830 (line "2012_20" in Table 3). These events might be more reliable and manifest actual seismicity of the North Atlantic. However, we retain in mind that longer windows also suppress true weak signals from low-magnitude events and many short signals are likely missed.

Table 4 lists the number of detections for ten stations in four frequency bands used for cross correlation and detection. The detection performance varies significantly from station to station and with frequency band. Station TORD has the highest number of detections in 2009 and 2012: 31,386 and 26,774, respectively. For AKASG, the number of detection is unexpectedly low considering its input to the REB cross correlation discussed in the previous Section. On average, four stations within North America have more detections than arrays in Eurasia, with TXAR being the most sensitive to signals from the northern part of the Atlantic Ocean.

Two lower frequency bands provide the largest portion of all detection at all stations except FINES and MKAR. We suggest that most of these detections are related to low-frequency and highly-coherent noise. These noise arrivals should be effectively eliminated by longer templates. At best stations, 20-second master signals reduce the number of detections by a factor of 10 to 50. Longer windows can be even more effective at higher frequencies. At FINES, the frequency band 2.0 Hz to 4.0 Hz gave the highest overall input because of the well-known ambient noise conditions. For MKAR, higher frequency detections are not usual but not excluded.

Table 4. Number of associated detections by ten IMS stations in four frequency bands, BP.

|  | AKASG | BRTR | FINES | GERES | ILAR | MKAR | PDAR | TORD | TXAR | YKA |
|---|---|---|---|---|---|---|---|---|---|---|
| 2009 | | | | | | | | | | |
| P0820 | 706 | 10723 | 1419 | 5006 | 13547 | 3977 | 7172 | 17212 | 13672 | 15488 |
| P1530 | 781 | 7333 | 1040 | 2825 | 4074 | 4893 | 4308 | 4704 | 9444 | 2870 |
| P2040 | 106 | 4750 | 2150 | 2083 | 4396 | 4079 | 2599 | 5876 | 5976 | 535 |
| P3060 | 22 | 1190 | 1077 | 2437 | 485 | 1101 | 2002 | 3594 | 1768 | 100 |
| Total | 1615 | 23996 | 5686 | 12351 | 22502 | 14050 | 16081 | 31386 | 30860 | 18993 |
| 2012 | | | | | | | | | | |
| P0820 | 1280 | 9635 | 733 | 4504 | 10017 | 2772 | 10003 | 12576 | 10091 | 11156 |
| P1530 | 491 | 6808 | 714 | 1379 | 4805 | 3990 | 4726 | 5667 | 7975 | 3172 |
| P2040 | 17 | 1715 | 1130 | 1464 | 4362 | 3321 | 2720 | 6227 | 4148 | 274 |
| P3060 | 6 | 31 | 200 | 1 | 127 | 213 | 744 | 2304 | 667 | 2 |
| Total | 1794 | 18189 | 2777 | 7348 | 19311 | 10296 | 18193 | 26774 | 22881 | 14604 |
| 2012_20 | | | | | | | | | | |
| P0820 | 104 | 1068 | 230 | 277 | 432 | 532 | 1473 | 1201 | 1383 | 364 |
| P1530 | 7 | 266 | 112 | 6 | 70 | 125 | 61 | 254 | 127 | 43 |
| P2040 | 0 | 86 | 64 | 78 | 144 | 76 | 18 | 101 | 77 | 5 |
| P3060 | 1 | 1 | 1 | 0 | 0 | 10 | 8 | 14 | 19 | 0 |
| Total | 112 | 1421 | 407 | 361 | 646 | 743 | 1560 | 1570 | 1606 | 412 |

In order to distinguish between valid and false detections we compare their principal properties. We analyze only those detections, which are associated with events in the 2009 XSEL and two 2012 XSELs as obtained with standard and long templates. Figure 13 shows that most of detections have small *SNR_CC* and even smaller standard SNR indicating their low reliability. For proved detections from the REB in Figure 10, the distribution of *SNR_CC* is practically exponential while the relevant distribution in Figure 13 has a sharp peak at 2.5. The exponential segment in these distributions starts at *SNR_CC*=5, with 333 and 330 detections having *SNR_CC*>6 in 2009 and 2012, respectively. All in all, the number cross correlation detections could be reduced by two orders of magnitude when a higher *SNR_CC* threshold is applied. This was the case for the aftershock sequence of the October 2011 earthquake (Bobrov *et al*., 2012a). The distribution of SNR peaks between 1 and 2 and such detections are definitely not good for the REB. At the same time, the level of CC in Figure 13 for regular templates is similar to those in Figure 11. Longer templates suppress CC estimates for weak signals and noise.

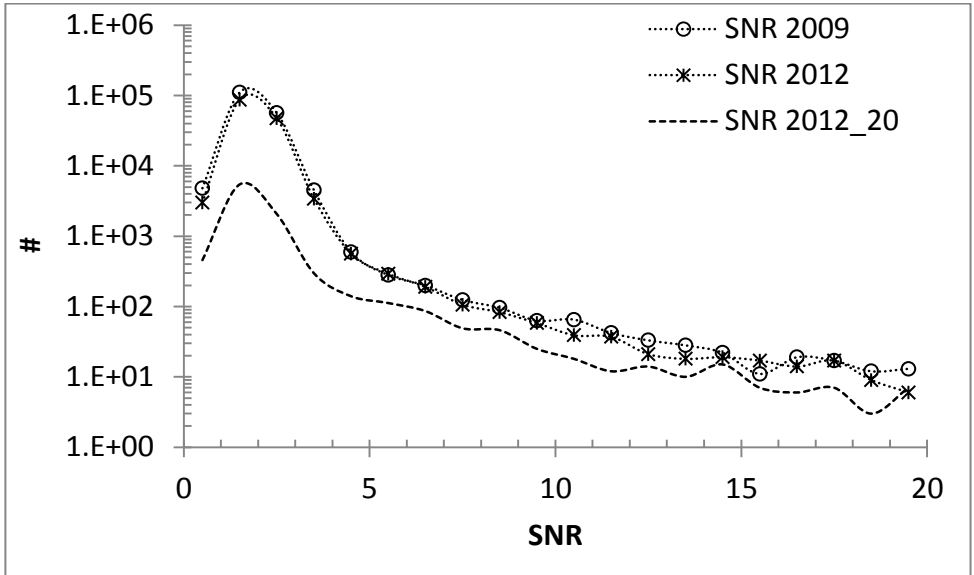

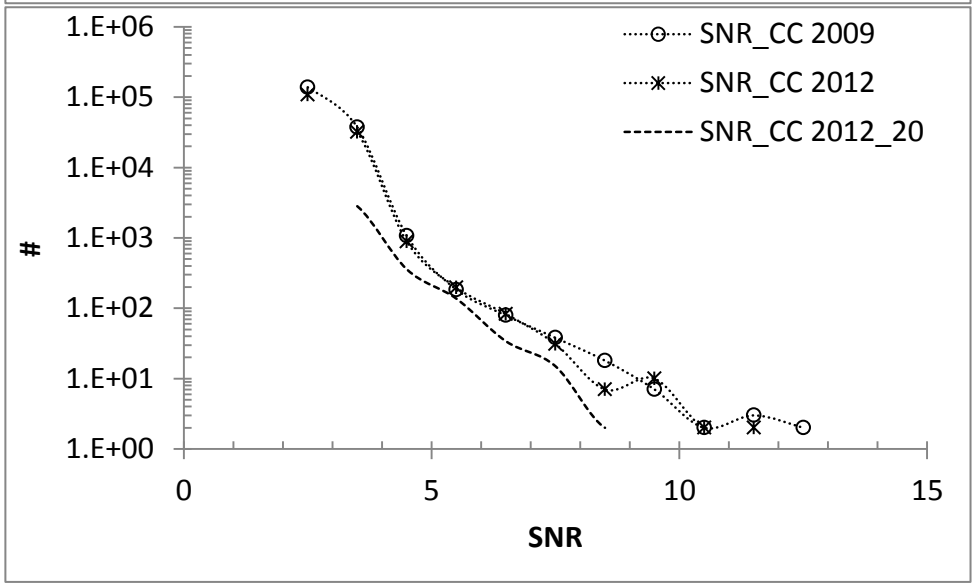

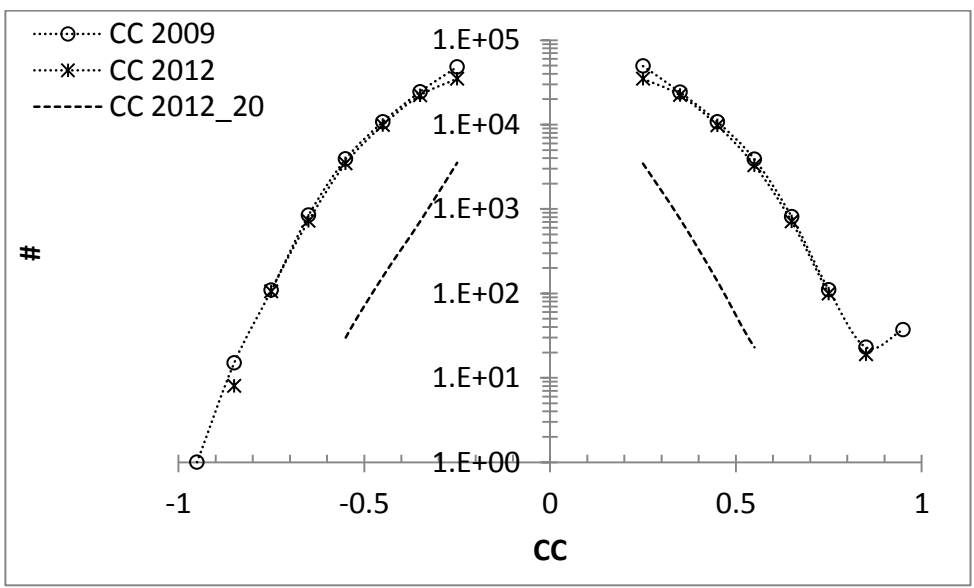

Figure 13. The overall frequency distribution of *SNR*, *SNR_CC*, and *CC* in three XSELs obtained in the first halves of 2009 and 2012. Dashed line represents the 2012 XSEL based on detections using 20-second-wide templates.

These observations provide strong evidence that the low-SNR signals are rather related to coherent noise or continuous low-amplitude signals from the Mid-Atlantic Ridge. In addition, the estimates of azimuth and slowness residuals in Figure 14 demonstrate almost uniform frequency distributions also indicating the coherent noise as the source of signals. It is especially important that all distribution have no prominent peak near 0, i.e. the signals with azimuths and slownesses different from those of the master event have the same chance to be associated. Therefore, the involved masters have a very limited azimuth and slowness resolution for low-SNR signals from the Mid-Atlantic Ridge.

The abundance of high CC detections is a challenge to the use of cross correlation for purposes of event building. The cross correlation properties of the built events might be helpful to distinguish between valid and invalid ones. Figure 15 depicts the frequency distributions of average and cumulative CCs for three studied XSELs. Most of the events have *CC_AV* values much lower than those in Figure 9 for the REB events with the highest *CC_AV* of 0.7. Three *CC_CU* curves do not demonstrate the same exponential roll-off as the observed for 931 REB events. Most of these XSEL events are unreliable and likely invalid.

All three *RM* frequency distributions in Figure 16 have peaks at -0.7. The corresponding IDC magnitude estimates for these XSELs peak near 3.8. This could be the detection threshold for the North Atlantic if not the input of coherent noise. The number of three station events shown in Figure 16 is by an order of magnitude larger than that of four station events. The events with five and more defining stations are more reliable and their total number does not differ between short (118 events) and long (91 event) template windows. In 2009, there were a few 9- and 10-station events, which were absent in 2012. It is likely that the events characterized by low *CC_AV* and the average *RM* below -0.5 deserve rejection. In this study, we intentionally put all thresholds low and opened the XSEL for invalid events. Based on a well prepared training set, which should include valid and invalid events and detections, standard classification algorithms may be used for removing invalid hypotheses.

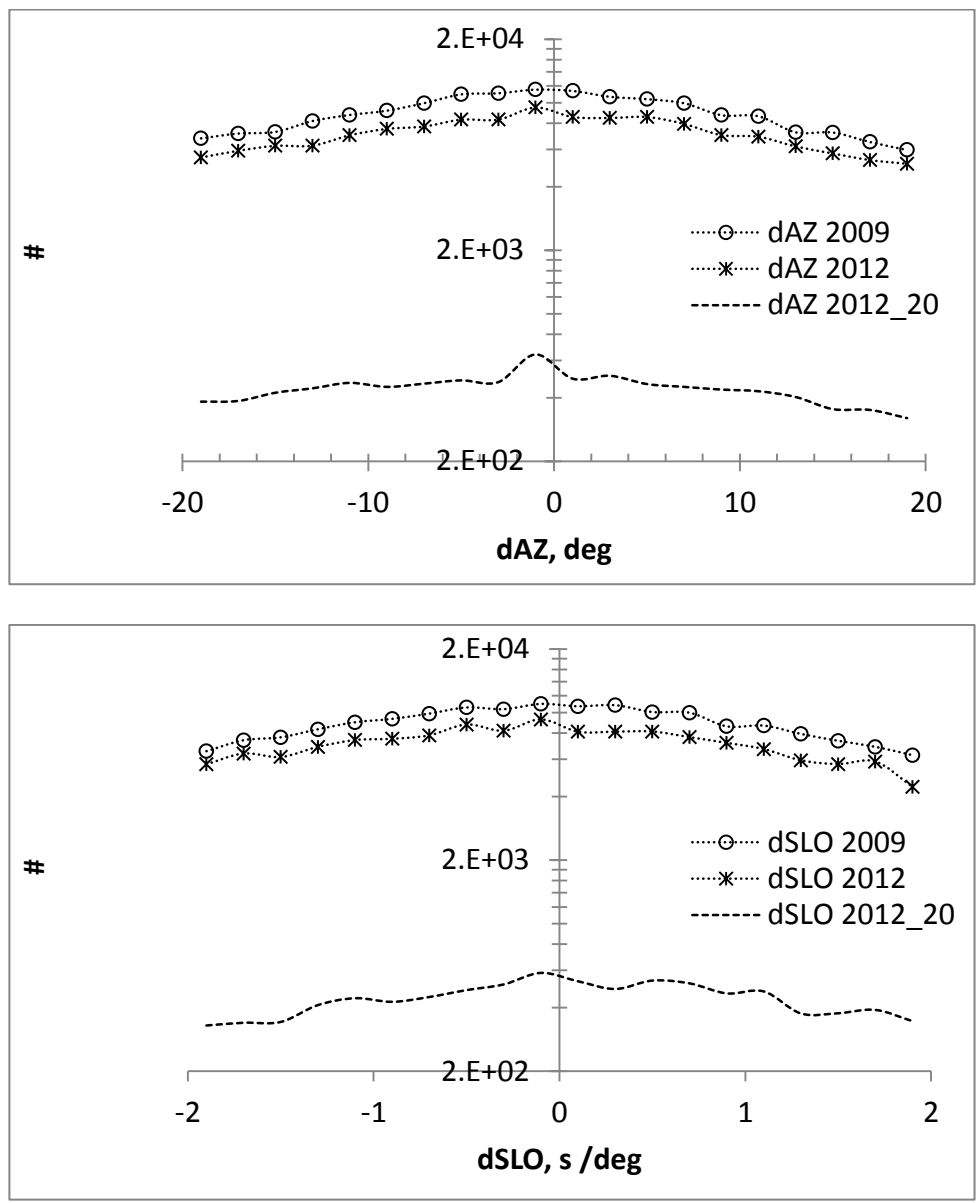

Figure 14. The overall frequency distribution of d*AZ* and *dSLO* in three XSELs obtained in the first halves of 2009 and 2012.

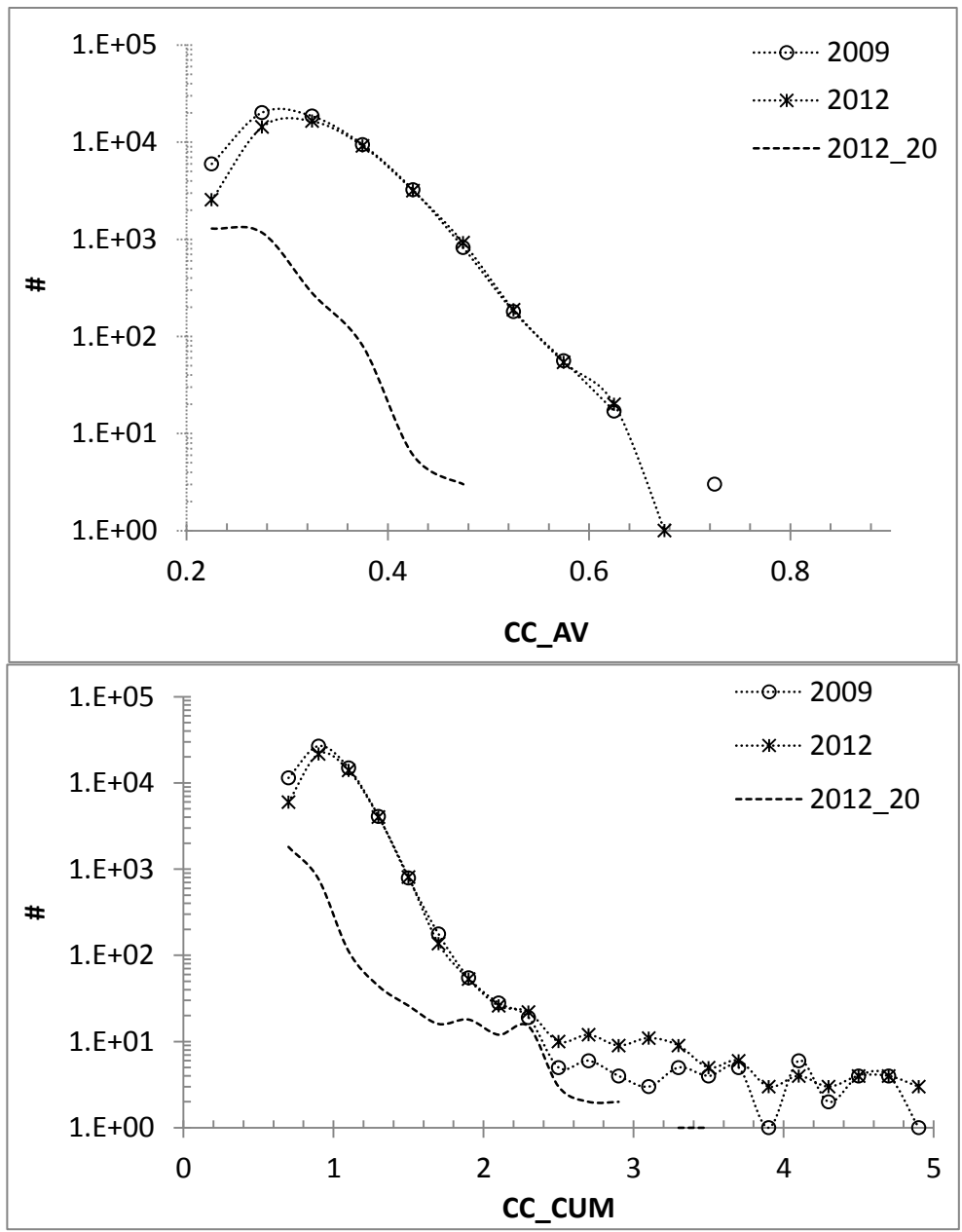

Figure 15. The frequency distribution of the average and cumulative cross correlation coefficient of three XSELs obtained in the first halves of 2009 and 2012.

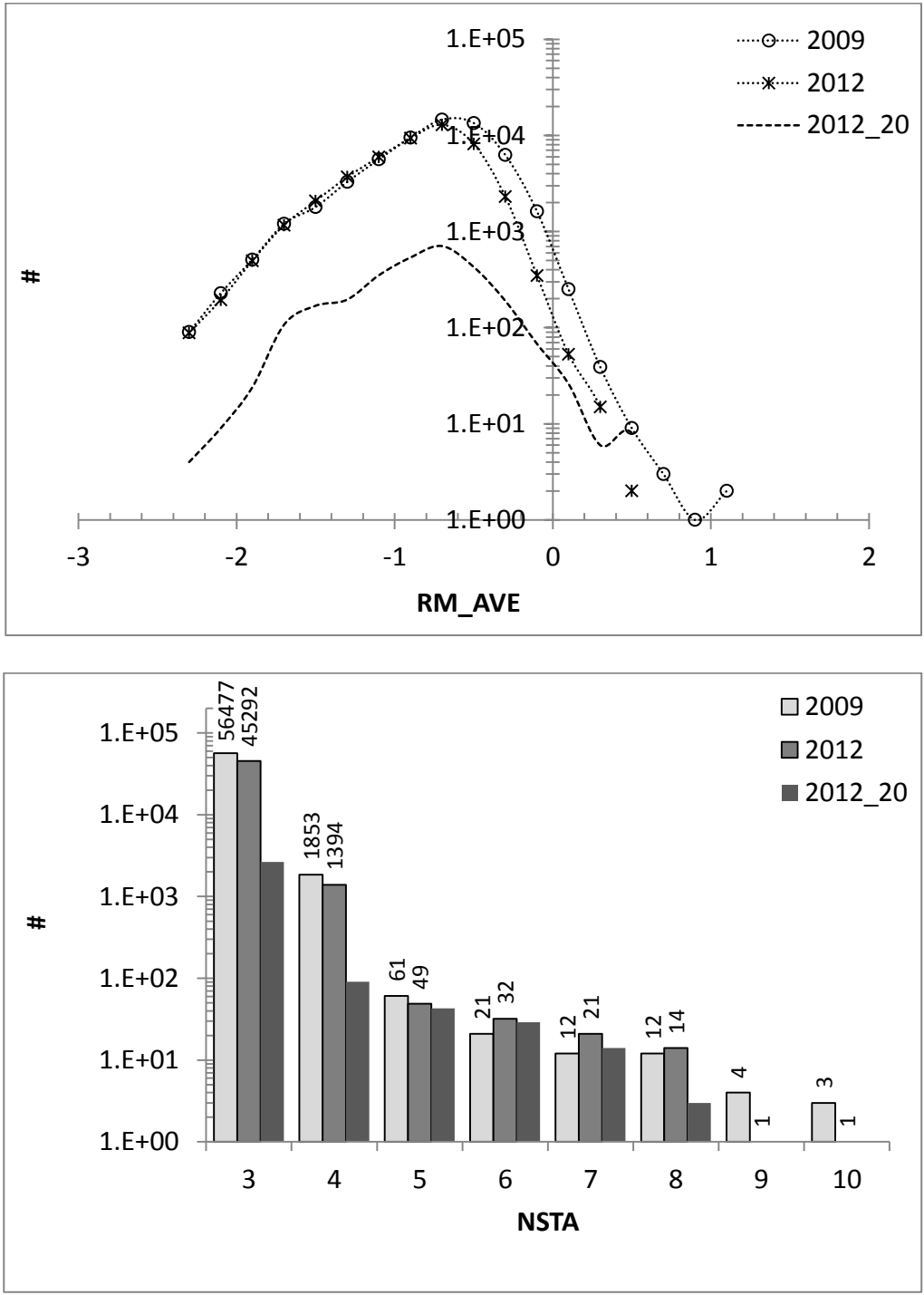

Figure 16. The frequency distribution of the average relative magnitude and the number of stations in the XSEL events obtained in the first halves of 2009 and 2012.

The quality of an XSEL event is defined not only by its averaged characteristics, but also by individual features of associated arrivals. We have estimated the azimuth, dAZ, and slowness, dSLO, residuals using the multichannel *CC* traces as well as the relative magnitude residual, d*RM*, which shows the dynamic consistency of involved arrivals. Figures 17 through 19 illustrate station dependence of these residuals and also the frequency distribution of the *SNR_CC* estimates, which can be used to select appropriate detections. On average, station ILAR shows larger *CC* estimates than those obtained at AKASG. It may be an indication of higher reliability of signals detected by ILAR and reduce the overall probability of hypotheses with AKASG as defining station. MKAR shows that the estimates of d*RM* and d*SLO* are distributed randomly with low-amplitude peaks. The peak of the slowness distribution is shifted by -0.5 s/deg from the central position. This bias is likely related to the non-planar propagation of the P-wave along the array. The velocity structure immediately beneath MKAR is inhomogeneous that leads to the difference between theoretical and empirical time delays at individual elements. This difference is expressed in a slightly shifted peak as estimated by the *f-k* analysis applied to the CC-traces. This is a well-known effect for array stations which is usually compensated by static corrections (Coyne *et al*., 2012). For longer templates, both distributions are sharper since the portion of noise detections is substantially reduced. The best quality indicator for cross-correlation detection is likely SNR_CC. The difference in detection quality between stations plays important role in phase association and event building. For example, Figure 19 shows that FINES produces less good SNR_CCs in relative terms than YKA. Therefore, YKA demonstrates a higher probability of quality detections and may be preferable in event building.

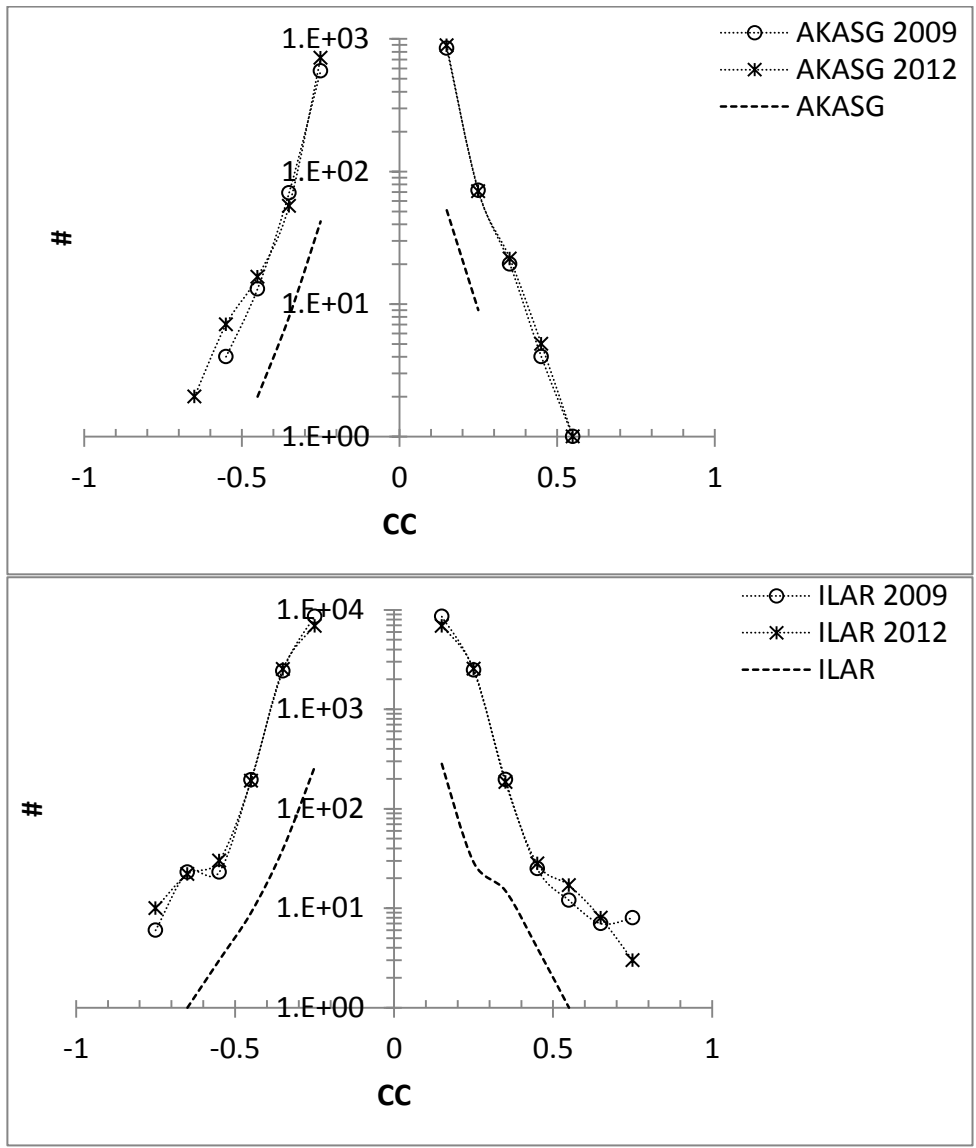

Figure 17. The frequency distribution of CC at stations AKASG and ILAR in three XSELs obtained in the first halves of 2009 and 2012.

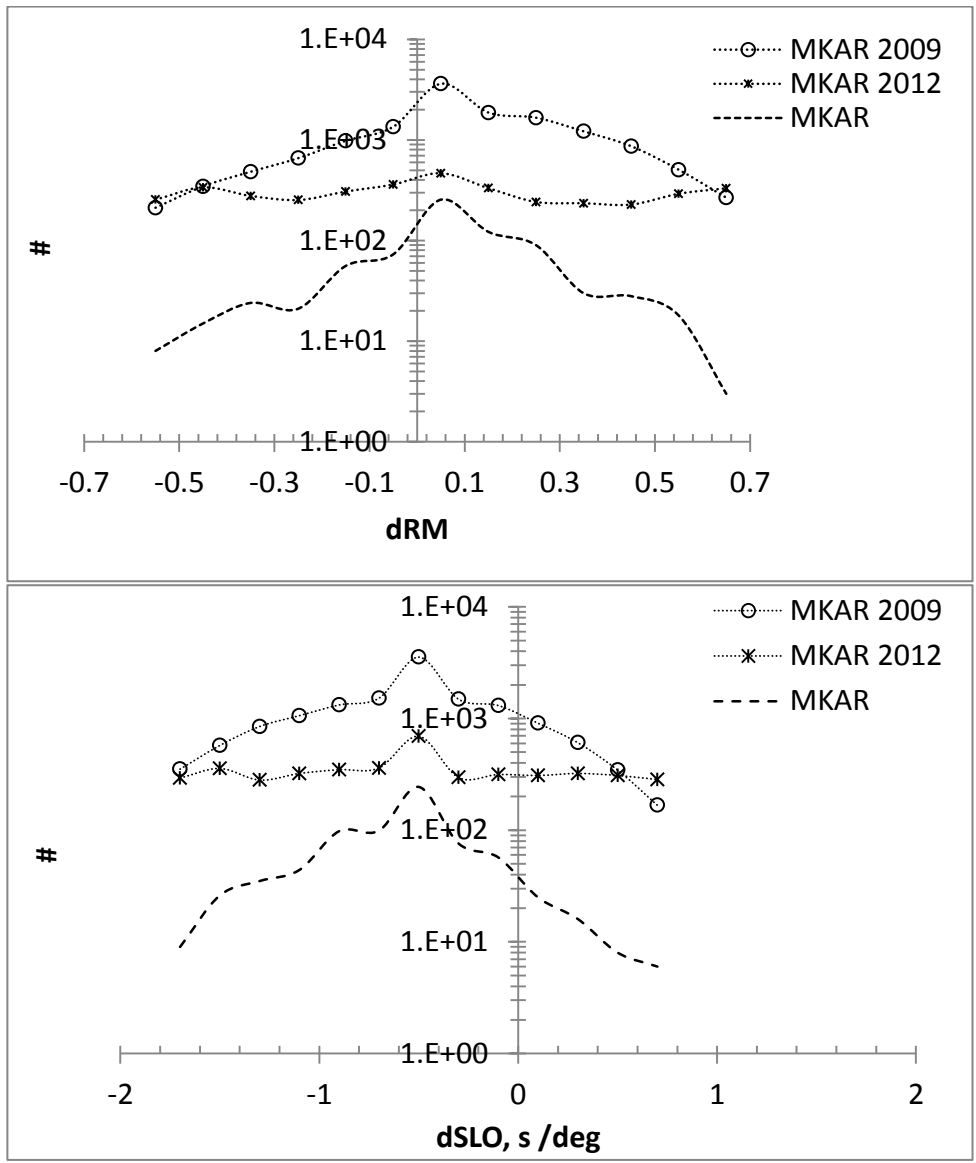

Figure 18. The frequency distribution of dRM and d*SLO* at station MKAR in three XSELs obtained in the first halves of 2009 and 2012.

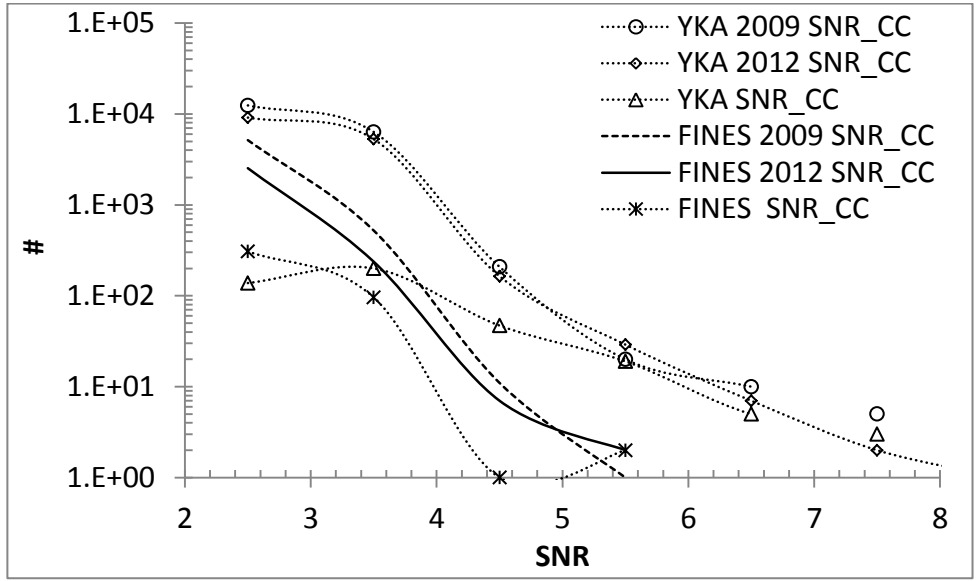

Figure 19. The frequency distribution of *SNR_CC* at station YKA and FINES in three XSELs obtained in the first halves of 2009 and 2012.

Here, we test the effect of azimuth, slowness, and relative magnitude residuals as well as CC and SNR_CC on the reliability of three-station XSEL hypotheses. Variable *SNR_CC* is likely the best for characterization of the quality of XSEL hypotheses. For 2009, there were 56,477 three-station hypotheses in the XSEL. There were 102 different triplets with most frequent sets: ILAR-TORD-TXAR (3068 hypotheses), TORD-TXAR-YKA (2859), and BRTR-TORD-TXAR (2817). Then we eliminated all triplets with at least one arrival having *SNR_CC*<3.0. There were only 569 (1%) hypotheses and 71 different triplets left with the most popular sets: ILAR-TORD-YKA (42), TORD-TXAR-YKA (40), and ILAR-TORD-TXAR (34). This observation is good evidence that we have put the detection threshold too low. A slight change to *SNR_CC*>3.0 produces a reasonable number of event hypotheses to be tested interactively.

In order to assess the portion of valid XSEL hypotheses machine learning can be applied. This procedure has been already tested on the example of the 2012 Sumatera aftershock sequence (Kitov *et al*., 2013b). We successfully found approximately 700 valid

XSEL hypotheses from the total number of ~2700 hypotheses applying a bootstrap aggregation (bagging) procedure for an ensemble of decision trees. We used the MATLAB TreeBagger construction for bagging. This result was confirmed by interactive review of valid and invalid XSEL hypotheses.

Following the procedure developed in (Kitov *et al.*, 2013) we have compiled a dataset containing two classes of arrivals: valid and invalid. The best way to select correct arrivals is analyst review of XSEL hypotheses. Since human resources needed for accurate and time-consuming review were not available for this specific purpose we have selected those XSEL events which had close origin times to the events reported in the ISC catalogue. The ISC is a major aggregator of regional seismological information and produces most accurate global catalogues, although with a two-year delay. The IDC is one of key contributors but may miss many events reported by regional networks.

For the first half of 2009, the not reviewed ISC catalogue for the northern part of the Atlantic Ocean includes 611 events (711 for the whole seismic region 32). The reviewed ISC catalogue has only 234 (323) events for the same period including 134 (231) events detected by the IDC. In both versions of the ISC catalogue, there are 73 events unique to the IDC. The difference between the reviewed and not reviewed ISC catalogues is mainly defined by the input of two agencies. The European Mediterranean Seismological Center (CSEM/EMSC) reported extra 194 events and the University of the Azores (PDA) added 157 events not confirmed by the ISC. These events belong to the Azores archipelago and the Gloria Fault. The master events located along the Mid-Atlantic Ridge do not cover these smaller earthquakes not detected at teleseismic distances.

Having origin times of 611 ISC events, we selected all XSEL event hypotheses within 40 s from the relevant ISC origin times. We use the origin time as a defining parameter since smaller events usually have poor locations and thus cannot be compared directly by their positions. The 40 s difference in origin times takes this mislocation in account. In total, there were 393 XSEL hypotheses within 40 s from at least one ISC event. However, we selected only 258 XSEL events for the training set. Generally, big earthquakes create coda-waves consisting of a larger number of reflected/refracted waves repeating the direct P-wave. Due to the similarity of signal shapes and propagation paths, waveform cross correlation effectively detects these later arrivals and produces virtual events with slightly different origin times. We rejected 135 XSEL hypotheses from 393 because they were based on the coda waves. The set of invalid events was created by random choice of ~3% (1859) of XSEL hypotheses far enough (600 s) from the ISC events.

In total, the class of valid events included 1031 arrivals and there were 5631 invalid arrivals. Using TreeBagger MATLAB application and this learning set, we have classified 177,520 arrivals in 48,443 XSEL hypotheses. There were 252 events defined by three or more valid arrivals. Figure 20 displays locations for these qualified XSEL events, which repeat the pattern of the Mid-Atlantic Ridge seismicity. To be included in the REB, all selected hypotheses have to be confirmed interactively. As in our previous studies, cross correlation approximately doubled the amount of REB events and reduced the detection threshold by 0.4 magnitude units.

## Discussion

We have investigated the level of cross correlation between signals generated by earthquakes in the Atlantic Ocean and recorded by IMS array stations at teleseismic distances. It has been found that signals from the opposite ends of the Mid-Atlantic Ridge have similar shapes (|CC| ~0.5). For arrivals with SNR>5, cross correlation coefficients may reach 0.7 and even more for closer events. For the weakest signals, cross correlation traces provide higher SNRs than those obtained from the original waveforms. This defines the advantage of cross correlation in detection.

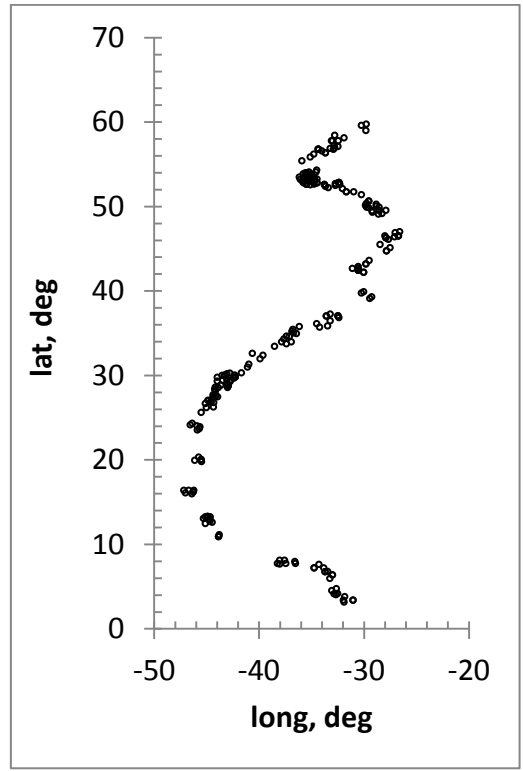

Figure 20. Locations of 252 XSEL hypotheses.

As a contributor to the International Seismological Centre, the IDC makes the REB available to the broader seismological community. The IMS has an advantage of global coverage with array stations, which provide higher signal enhancement using cross correlation than that for standard beam forming. When applied to IMS seismic data, cross correlation allows finding more qualified REB events that improves the completeness of the ISC catalogue and the quality of the ISC Bulletin.

In this study, we demonstrated the possibility of substantial improvements in the completeness of the IDC bulletin. The improvement of the IDC detection and monitoring threshold is most valuable for remote regions without regional networks. Having analysed the vast area of the Mid-Atlantic Ridge one can transport all general findings to other seismic regions poorly covered by seismic networks.

For the Atlantic Ocean, the gain from cross correlation is hindered by practically permanent generation of coherent noise associated with active tectonic and volcanic activity along the Mid-Atlantic Ridge. The cross correlation detector tuned to continental aftershock sequences finds enormous number of weak signals, which are physically valid but cannot be associated with valid REB events. With increasing template length, these spurious detections are better suppressed. Unfortunately, many valid but weak signals are also eliminated. The high level of coherent noise from the ridge also affects the beam forming technique: the detection threshold for routine IDC processing is high and the number of REB events in region 32 is relatively low. All in all, the relative gain in detection provided by cross correlation in the Atlantic Ocean is as high as in different regions.

A larger portion of detections obtained using cross correlation with waveform templates from 60 master events have *SNR_CC* between 2.5 and 4.0. Our analysis showed their low reliability and limited usefulness for the XSEL. In addition, the estimates of relative magnitude, azimuth, and slowness residuals demonstrate uniform distribution also revealing coherent noise as the source of spurious arrivals. Many events are built by low SNR_CC detections with relatively high CC. These events are unreliable and deserving rejection. The CC and SNR_CC thresholds in this study were very low and opened the XSEL for invalid events. Machine learning and classification algorithms effectively remove these invalid hypotheses.

## Acknowledgements

This publication has been produced with the assistance of the European Union, EU Council Decision 2010/CFSP of 26 July 2010.

## Disclaimer

The contents of this publication are the sole responsibility of the authors and can in no way be taken to reflect the views of the European Union and the CTBTO Preparatory Commission.